# The Parallel Ionizing Emissivity Survey (PIE). I. Survey Design and Selection of Candidate Lyman Continuum Leakers at $3.1 < z < 3.5$

Alexander Beckett[1], Marc Rafelski[1,2], Claudia Scarlata[3], Wanjia Hu[3], Keunho Kim[4], Ilias Goovaerts[1], Matthew A. Malkan[5], Wayne Webb[5], Harry Teplitz[4], Matthew Hayes[6], Vihang Mehta[4], Anahita Alavi[4], Andrew J. Bunker[7], Annalisa Citro[3], Nimish Hathi[1], Alaina Henry[1], Alexandra Le Reste[3], Alessia Moretti[8], Michael J. Rutkowski[9], Maxime Trebitsch[10], and Anita Zanella[11]

[1] Space Telescope Science Institute, 3700 San Martin Drive, Baltimore, MD 21218, USA; abeckett@stsci.edu
[2] Department of Physics and Astronomy, Johns Hopkins University, Baltimore, MD 21218, USA
[3] Minnesota Institute for Astrophysics, University of Minnesota, Minneapolis, MN 55455, USA
[4] IPAC, Mail Code 314-6, California Institute of Technology, 1200 E. California Boulevard, Pasadena, CA 91125, USA
[5] UCLA, Department of Physics and Astronomy, Los Angeles, CA 90095, USA
[6] Stockholm University, Department of Astronomy and Oskar Klein Centre for Cosmoparticle Physics, SE-10691 Stockholm, Sweden
[7] Department of Physics, University of Oxford, Denys Wilkinson Building, Keble Road, Oxford OX1 3RH, UK
[8] INAF—Osservatorio Astronomico di Padova, Vicolo dell'osservatorio 5, 35122, Padova, Italy
[9] Minnesota State University-Mankato, Dept. of Physics & Astronomy, Trafton Science Center North 141, Mankato, MN 56001, USA
[10] LERMA, Sorbonne Université, Observatoire de Paris, PSL Research University, CNRS, 75014 Paris, France
[11] INAF—Osservatorio di Astrofisica e Scienza dello Spazio di Bologna, Via Gobetti 93/3, 40129, Bologna, Italy

## Abstract

We present the survey design and initial results from the Parallel Ionizing Emissivity (PIE) survey. PIE is a large Hubble Space Telescope survey designed to detect Lyman continuum (LyC) emitting galaxies at $3.1 < z < 3.5$ and stack their images in order to measure average LyC escape fractions as a function of galaxy properties. PIE has imaged 37 independent fields in three filters (F336W, F625W, and F814W), of which 18 are observed with a fourth band (F475W) from the accompanying PIE+ program. We use photometric colors to select candidate Lyman break galaxies (LBGs) at $3.1 < z < 3.5$, which can be followed up using ground-based spectrographs to confirm their redshifts. Unlike previous surveys, we use many independent fields to remove biases caused by correlated absorption in the intergalactic medium (IGM). In this paper, we describe the survey design, photometric measurements, and the use of mock galaxy samples to optimize our color selection. With three filters, we can select a galaxy sample of which ≈90% are LBGs and over 30% lie in the $3.1 < z < 3.5$ range for which we can detect uncontaminated LyC emission in F336W. We also use mock IGM sight lines to measure the expected transmission of the IGM, which will allow us to determine escape fractions from our stacked galaxies. We color-select ≈1400 galaxies, and predict that this includes ≈80 LyC-emitting galaxies and ≈500 that we can use in stacking. Finally, we present the Keck/LRIS spectrum of a galaxy at $z \approx 2.99$, demonstrating that we can confirm the redshifts of $z \sim 3$ galaxies from the ground.

## 1. Introduction

The reionization of the intergalactic medium (IGM), taking place mostly between redshifts of ≈10 and ≈5 (e.g., R. H. Becker et al. 2001; A. J. Pahl et al. 2020; Y.-H. Lin et al. 2024) is the last major phase transition in cosmic history, affecting how galaxies formed and evolved. Galaxies likely drove reionization, but we do not yet have a clear understanding of what properties make a galaxy an effective ionizing source, and therefore which galaxies emitted most of the ionizing photons (e.g., S. L. Finkelstein et al. 2019; R. P. Naidu et al. 2020; J. Witstok et al. 2025). This requires a direct measurement of $f_{esc}$, the escape fraction of hydrogen-ionizing Lyman continuum (LyC) radiation from galaxies.

Unfortunately, due to high attenuation from the IGM, direct observations of LyC emission at $z \gtrsim 4.5$ are virtually impossible (A. K. Inoue et al. 2014). We therefore need to identify indirect indicators of LyC emission at lower redshifts, where we can measure LyC alongside other galaxy properties. These measurements would allow us to infer LyC emission from observables during the epoch of reionization (EoR).

Measuring $f_{esc}$ presents an observational challenge at all redshifts. At low redshifts, this requires observations of the far-UV, which does not reach ground-based telescopes. At higher redshifts, the intrinsically faint LyC is affected by cosmological dimming and strongly attenuated by the IGM, requiring deep observations. For a long time, most studies failed to detect ionizing radiation from star-forming galaxies (e.g., C. Leitherer et al. 1995; B. Siana et al. 2007; C. R. Bridge et al. 2010; B. Siana et al. 2015), although some galaxies and/or stacks were found to leak LyC flux (e.g., C. C. Steidel et al. 2001). With more recent measurements, we now have directly measured $f_{esc}$ for dozens of galaxies (e.g., Y. I. Izotov et al. 2016b, 2018a, 2018b; C. C. Steidel et al. 2018; S. R. Flury et al. 2022; J. Kerutt et al. 2024).

Many of these are low-redshift ($z \approx 0.3$) galaxies known as "Green Peas" (C. Cardamone et al. 2009): young, compact, highly ionized, highly star-forming galaxies similar to those expected in the early Universe (R. Amorín et al. 2012; A. E. Jaskot & M. S. Oey 2013). Several diagnostics are found to trace LyC emission in these galaxies, including the peak separation of the Ly$\alpha$ emission line (e.g A. Verhamme et al. 2017), as well as the ionization state and properties such as specific star formation rate and star formation surface density (S. R. Flury et al. 2022). Many of these low-$z$ leakers show strong Ly$\alpha$ emission, and are often found to have $f_{\rm esc} \gtrsim 5\%$ and up to 70% (e.g., Y. I. Izotov et al. 2016a, 2016b, 2018a, 2018b), although the presence of Ly$\alpha$ requires $f_{\rm esc} < 100\%$. However, several galaxies have been found with strong Ly$\alpha$ that are not strong LyC leakers (e.g., A. Citro et al. 2025), and LyC emission has been seen in galaxies with very different properties than the green peas (e.g., N. Roy et al. 2024).

It is also somewhat unclear to what extent LyC leakers in the EoR share the properties of low-redshift leakers such as the green pea galaxies. These calibrators are likely to evolve with redshift (e.g., A. E. Jaskot et al. 2024), so need to be calibrated on sources closer in time to the epoch of reionization. LyC-emitting galaxies have been found at $z > 3$ (e.g., A. E. Shapley et al. 2016; E. Vanzella et al. 2016, 2018; T. J. Fletcher et al. 2019; R. Marques-Chaves et al. 2021; L. J. Prichard et al. 2022; U. Meštrić et al. 2025), allowing tests of diagnostics such as Ly$\alpha$ shape and UV luminosity (e.g., A. J. Pahl et al. 2023; A. Pahl et al. 2024). However, some of these results suggest differences in the relationships between these properties and LyC emission (e.g J. Kerutt et al. 2024). These studies have not yet revealed a direct proxy for $f_{\rm esc}$ that could be used to infer LyC emission during reionization.

The measured LyC flux at these redshifts depends on both $f_{\rm esc}$ and the transmission of the IGM ($T_{\rm IGM}$). $T_{\rm IGM}$ is impractical to measure for any individual sight line, and is correlated on scales of several Mpc due to the large-scale cosmic structure (e.g., C. M. Scarlata et al. 2025). Therefore, single-galaxy detections and small samples taken from individual deep fields (such as the UV fields in COSMOS, L. J. Prichard et al. 2022) are likely to provide biased measurements or retain large uncertainties due to the unknown $T_{\rm IGM}$. In order to reduce these uncertainties and provide a large, unbiased sample from which to measure $f_{\rm esc}$, we require a large number of fields that are well-separated to ensure no correlations in their $T_{\rm IGM}$.

The Parallel Ionizing Emissivity (PIE) program is an ongoing effort to observe dozens of independent fields with the Hubble Space Telescope (HST), in order to identify a large sample of $z > 3$ galaxies free from these biases, and directly constrain their LyC output. In this paper, we describe the survey design and present results from the HST data, which also include additional imaging from the PIE+ program. Section 2 describes the survey strategy and the observations. Section 3 details the reduction of the HST data in order to produce our science images. In Section 4 we describe the simulated galaxies we use to validate our photometric measurements and color selection, which are themselves presented in Sections 5 and 6, respectively. We show an early example from our spectroscopic observations in Section 7, and our main conclusions are summarized in Section 8.

Throughout this work, we assume a $\Lambda$ cold dark matter cosmology with $H_0 = 67.4\ \rm km\ s^{-1}\ Mpc^{-1}$, and $\Omega_m = 0.315$, as found by Planck Collaboration (2020).

## 2. Survey Design

### 2.1. Program Overview

The PIE survey is designed to measure the properties of galaxies across an unbiased sample at $3.1 < z < 3.5$, with the aim of measuring $f_{\rm esc}$ as a function of galaxy properties. This is close to the highest redshift at which LyC radiation can be detected; the IGM becomes almost entirely opaque by $z \approx 4$ (e.g., A. K. Inoue et al. 2014). By measuring correlations between $f_{\rm esc}$ and galaxy properties at these redshifts, the PIE data can bridge the gap between low-$z$ galaxies such as the "green peas" and the galaxies that reionized the Universe at $z > 5$. We will combine the HST photometric and morphological data with ground-based spectroscopy, in order to test several diagnostics that could be used to infer $f_{\rm esc}$ during the epoch of reionization itself, one of the key goals of JWST.

The PIE program is a three-band survey using the Wide Field Camera 3 (WFC3) on HST (e.g., M. Marinelli & L. Dressel 2024). This is a pure-parallel program, initially targeting 65 random fields across the sky in order to avoid any biases caused by correlations in the opacity of the IGM (ID: 17147, PI: Scarlata). We have observed using 59 opportunities,[12] obtaining data for 54 unique fields (since a few of the opportunities cover the same parallel field). Our observing strategy, using F336W to detect LyC emission, sets our target redshift range: we detect pure LyC emission only at $z > 3.1$, and the opacity of the IGM obviates LyC detection above $z \approx 3.5$ (see Section 4.3).

We use color cuts (described in Section 6) to select high-redshift galaxies, based primarily on their Lyman break. Although we target the $3.1 < z < 3.5$ redshift range, three-band selection cannot reliably determine whether galaxies lie in this range, instead only allowing us to select $2.7 \lesssim z \lesssim 4.0$ Lyman break galaxies (LBGs). These targets will be followed-up with spectroscopic observations, both from ground-based spectrographs such as Keck/LRIS and from JWST, enabling redshifts to be confirmed and emission-line fluxes to be measured.

With spectroscopic and photometric data, we will be able to stack galaxies with different properties to measure $f_{\rm esc}$, measure morphologies of LyC emission, as well as compare optical morphologies of leakers and non-leakers. We will then be able to discuss how $f_{\rm esc}$ varies with changes in these galaxy properties. In this work, we focus on the HST data and the selection of targets for spectroscopic follow-up. We are still in the very early stages of our spectroscopic campaign, so we present only a brief example here (in Section 7). The final spectra and analysis thereof will be discussed in future papers.

### 2.2. HST Observing Strategy

The structure of our observations depends on the structure of visits and orbits chosen by the various "prime" programs alongside which we observed. For most of our parallel opportunities, we observe for one orbit in the optical, spending half an orbit on each of the F625W and F814W bands in order to allow color selection of LBGs. This time is usually split into three exposures for each of these bands to allow for reliable removal of cosmic rays. However, some fields with longer opportunities are observed for a full orbit in each of these filters, split into three longer exposures.

[12] A parallel opportunity is a set of visits for which the prime observations are from a single GO program covering a single target.

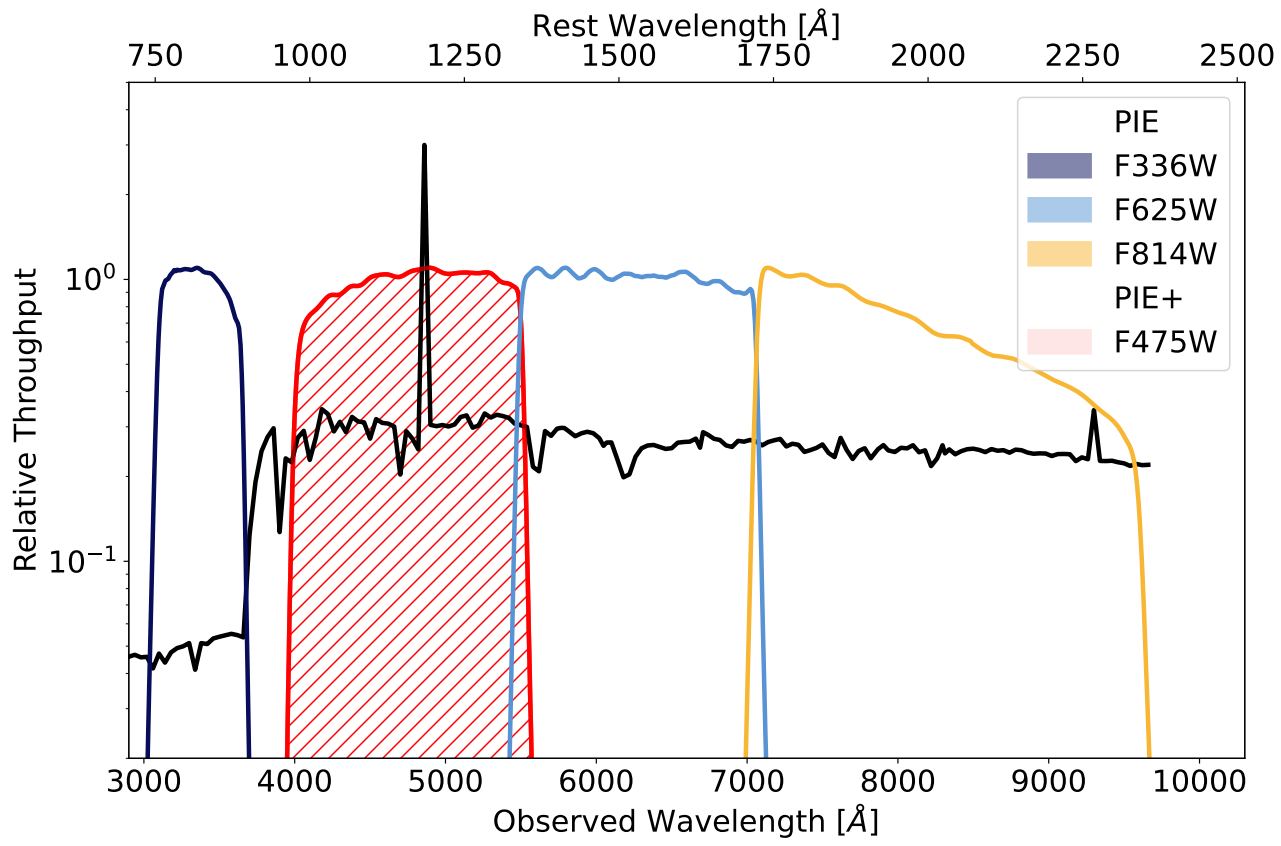

**Figure 1.** The filters used in this work. The primary PIE survey covers LyC emission in F336W, and measures optical colors using F625W and F814W for 37 fields. PIE+ adds F475W for 18 of these fields, which helps constrain the redshift. We show an illustrative spectrum with the Lyman break and Ly$\alpha$ emission clearly visible.

The remainder of each opportunity is used to maximize our depth in the F336W band and, hence, the number of galaxies for which we can detect LyC emission. Our fields therefore have a variety of exposure times in F336W, varying from $\approx$3000 to $\gtrsim$30,000 s. For our deepest fields ($\gtrsim$16,000 s), we use full-orbit exposures, reducing the noise from post-flash and readout. For most other fields, we use half-orbit exposures to ensure we have enough exposures to reliably remove cosmic rays and satellite trails.

In addition to the three-band observations that make up the PIE survey, 25 of our fields also feature observations from the PIE+ program (ID: 17518, PI: Beckett), 18 of these in F475W. Using this fourth filter, PIE+ allows for a wider range of color cuts to better identify high-redshift galaxies and, hence, increase the fraction of $3.1 < z < 3.5$ galaxies in our spectroscopic sample. Every PIE+ observation consists of three exposures of 500 s each, with a small dither to maximize resolution and mitigate the impact of hot pixels. These filters are shown in Figure 1, illustrating how the Lyman break is used to select galaxies at these redshifts.

We need successful observations in all three of the PIE filters to apply a color selection in each field, with three exposures in each band in order to remove cosmic rays. Unfortunately, due to issues affecting the HST gyros during the time period covering these observations, only 32 of our 54 observed fields have this level of usable data in all three bands. The other fields have at least one band for which the data are affected by tracks, or are missing entirely due to the telescope entering "safe mode." Although the PIE+ program was able to replace the affected optical data for five fields (instead of adding F475W), the others remain unusable. As the PIE+ data was taken in the cycle after the first PIE observations, this was much less affected, although there were four additional PIE+ observations that did not produce usable data.

Overall, our HST imaging allows us to perform color selection in 37 of our fields, of which 18 have all four PIE and PIE+ bands and 19 have only three bands of imaging. These fields are summarized in Table 1. Where some exposures are affected yet there is usable data for all three PIE bands, only the usable exposure time is listed here. Even for the fields that could not be used, the successful imaging could still be of interest to the community, so details of these observations are provided in Appendix A (Table 6). The field locations are also shown in Figure 2. We divide the fields by exposure time in F336W into "shallow" (<2 hr), "medium" (2–4 hr), and "deep" fields (>4 hr). The image depths listed in Table 1 are estimated by running SourceExtractor (E. Bertin & S. Arnouts 1996) on each of the science drizzles (using a similar process to that described in Section 5), and taking the median magnitude of sources with total signal-to-noise ratio (S/N) between 4.8 and 5.2.

We also note that two of our fields, with complete observations allowing for color selection, lie in close proximity to local galaxies (specifically, fields 41 and 51 lie close to Sextans B and NGC 4383, respectively). The large number of stellar sources in these fields makes efficient targeting of LBGs impossible. We do show these fields in Table 1 and Figure 2, as they are still useful for measuring completeness and other similar tests. However, we do not plan to follow these up with spectroscopy, leaving 35 fields that we will target with our spectroscopic survey.

## 3. Data Reduction

As the HST data that makes up the PIE survey is obtained through parallel observations, the exposures are not dithered. This requires some customized data reduction in order to minimize the impacts of hot pixels and cosmic rays, and obtain the best-quality images possible with our data. A similar procedure has been utilized to analyze WFC3 data in several other studies (e.g., L. J. Prichard et al. 2022; M. Revalski et al. 2023; X. Wang et al. 2025), but we describe the process in full below. We apply these same custom routines to the PIE+ data, although the dithering allows us to remove many of these artifacts, so the improvement in quality is smaller.

Our customized code first flags hot pixels using a spatially varying threshold. These are due to damage to the detector and, hence, are spread randomly across the image. Due to imperfect charge transfer efficiency (CTE), hot pixels are likely to be less prominent farther from the readout, so we need to use a varying threshold that drops with increasing distance from the readout in order to reliably flag these artifacts. We first measure the number of hot pixels in the 50 rows closest to the readout (so those least affected by CTE). For each subsequent set of 50 rows, we adjust the threshold such that each set of rows contains a similar number of hot pixels. This process is described in more detail in L. J. Prichard et al. (2022).

Next, we mitigate any offset in the background levels of our images caused by the amplifier response. We use PHOTUTILS (L. Bradley et al. 2024) to detect and mask sources in each image, then measure the background level in the unmasked region covered by each amplifier, and adjust this level to ensure it is consistent across each frame.

For optimally clean images, we must not only remove cosmic rays, but also the effects of cosmic rays during readout (ROCRs). We produce an image for each filter by combining the exposures using ASTRODRIZZLE, part of the DRIZZLEPAC software suite (S. Gonzaga et al. 2012; S. L. Hoffmann et al. 2021). This flags cosmic rays based on the differences between the exposures. As ROCRs affect the images closer to the readout than their location in the image would suggest, the CTE overcorrects, causing negative "divots." These are removed by flagging any pixels within 5 pixels of a cosmic ray that are more than 2.75$\sigma$ below the background level.

A more thorough detection of cosmic rays is then applied using LACOSMIC (P. G. van Dokkum et al. 2012). Finally, we

**Table 1**
Locations and Usable HST Depths and Exposure Times for Each PIE Field

| Field | R.A. | Decl. | Category | F336W | | F475W | | F625W | | F814W | |
|---|---|---|---|---|---|---|---|---|---|---|---|
| | | | | Exp (s) | $1\sigma$ Depth (mag) | Exp (s) | $5\sigma$ Depth (mag) | Exp (s) | $5\sigma$ Depth (mag) | Exp (s) | $5\sigma$ Depth (mag) |
| | hms | dms | | | | | | | | | |
| 01 | 10:37:09.642 | +37:09:30.66 | medium | 8209 | 29.3 | 1500 | 26.9 | 1150 | 26.5 | 1150 | 26.2 |
| 02 | 12:56:31.700 | −05:45:23.00 | shallow | 4142 | 28.6 | ⋯ | ⋯ | 1150 | 26.4 | 1120 | 26.0 |
| 04 | 13:43:11.505 | −00:51:27.41 | deep | 16837 | 29.5 | ⋯ | ⋯ | 2043 | 27.0 | 2046 | 26.7 |
| 05 | 09:56:21.151 | +28:48:06.93 | medium | 10477 | 29.3 | 1500 | 26.9 | 2276 | 27.1 | 2274 | 26.7 |
| 06 | 10:42:05.548 | +18:20:52.16 | shallow | 6259 | 29.1 | 1500 | 26.8 | 1150 | 26.6 | 1150 | 26.2 |
| 07 | 14:34:55.651 | +20:10:06.92 | medium | 8156 | 29.2 | ⋯ | ⋯ | 1150 | 26.6 | 1130 | 26.2 |
| 08 | 10:21:22.942 | +18:03:46.79 | deep | 15611 | 29.6 | 3000 | 27.2 | 2070 | 27.1 | 2370 | 26.8 |
| 09 | 08:37:01.250 | +19:18:39.05 | shallow | 5906 | 28.8 | 1500 | 26.8 | 1150 | 26.5 | 1150 | 26.1 |
| 12 | 09:49:27.882 | +48:30:50.28 | shallow | 6513 | 29.1 | ⋯ | ⋯ | 1158 | 26.6 | 1500 | 26.4 |
| 14 | 09:33:37.868 | +55:10:12.63 | deep | 15560 | 29.4 | 1500 | 27.0 | 1114 | 26.5 | 2280 | 26.5 |
| 15 | 12:47:37.246 | +58:19:56.88 | shallow | 3490 | 28.6 | ⋯ | ⋯ | 1101 | 26.4 | 1500 | 26.4 |
| 16 | 15:27:01.362 | −23:31:45.17 | deep | 27779 | 29.8 | ⋯ | ⋯ | 4788 | 27.4 | 2400 | 26.8 |
| 17 | 11:45:12.397 | +62:03:40.81 | medium | 7880 | 29.5 | 1500 | 26.8 | 1047 | 26.8 | 1152 | 26.2 |
| 22 | 22:29:31.974 | +27:30:54.07 | medium | 8590 | 29.2 | 1500 | 27.1 | 1150 | 26.6 | 1130 | 26.2 |
| 23 | 14:27:19.886 | +26:28:02.71 | shallow | 6763 | 29.1 | ⋯ | ⋯ | 1170 | 26.5 | 1170 | 26.1 |
| 25 | 13:40:12.756 | +54:43:38.30 | medium | 9470 | 29.2 | 1500 | 26.9 | 1200 | 26.6 | 1200 | 26.1 |
| 26 | 15:44:17.383 | +27:38:22.94 | medium | 9273 | 29.1 | ⋯ | ⋯ | 1170 | 26.5 | 1170 | 26.1 |
| 27 | 16:37:57.174 | +33:44:46.39 | medium | 8801 | 29.3 | ⋯ | ⋯ | 1150 | 26.6 | 1120 | 26.1 |
| 29 | 10:05:38.081 | +60:27:36.66 | shallow | 5845 | 28.8 | ⋯ | ⋯ | 1150 | 26.5 | 1130 | 26.1 |
| 31 | 09:27:42.579 | +30:47:27.31 | medium | 8140 | 29.1 | 1500 | 26.9 | 1120 | 26.4 | 1130 | 26.0 |
| 32 | 11:02:01.208 | +51:32:19.40 | deep | 16370 | 29.6 | 1500 | 27.0 | 2004 | 27.0 | 2571 | 26.9 |
| 33 | 12:02:35.358 | +55:08:17.80 | medium | 7675 | 29.2 | 1500 | 27.0 | 1145 | 26.5 | 1090 | 26.1 |
| 34 | 10:16:54.722 | +47:06:17.32 | deep | 21256 | 29.5 | 1500 | 27.0 | 1977 | 26.9 | 1977 | 26.5 |
| 35 | 09:57:10.764 | +28:51:35.55 | deep | 21178 | 29.4 | 3000 | 27.3 | 2300 | 26.8 | 2265 | 26.4 |
| 36 | 20:43:53.591 | −10:39:52.54 | medium | 13974 | 29.4 | 1500 | 26.8 | 2394 | 27.1 | 2388 | 26.7 |
| 37 | 10:11:22.926 | −04:39:54.77 | deep | 17634 | 29.5 | 1500 | 26.8 | 2082 | 26.9 | 2382 | 26.7 |
| 38 | 02:12:23.782 | +00:55:28.216 | deep | 15546 | 29.6 | ⋯ | ⋯ | 1500 | 26.7 | 2382 | 26.8 |
| 41 | 12:25:23.770 | +16:26:30.24 | medium | 8231 | 29.1 | ⋯ | ⋯ | 1150 | 26.4 | 1150 | 26.0 |
| 42 | 09:31:52.832 | +32:00:54.65 | medium | 8155 | 29.1 | ⋯ | ⋯ | 1145 | 26.5 | 1130 | 26.1 |
| 43 | 12:20:56.068 | +17:22:05.82 | medium | 8231 | 29.2 | ⋯ | ⋯ | 1150 | 26.5 | 1150 | 26.1 |
| 44 | 09:56:24.891 | +28:53:01.56 | medium | 9080 | 29.3 | ⋯ | ⋯ | 2270 | 27.0 | 2260 | 26.7 |
| 45 | 09:33:48.771 | +51:14:37.96 | medium | 9760 | 29.4 | ⋯ | ⋯ | 980 | 26.4 | 980 | 26.0 |
| 46 | 09:56:40.324 | +17:35:30.18 | shallow | 4000 | 28.7 | ⋯ | ⋯ | 2259 | 26.9 | 1500 | 26.2 |
| 47 | 09:09:09.284 | +39:34:25.88 | medium | 8180 | 29.2 | 1500 | 26.9 | 1020 | 26.4 | 1300 | 26.2 |
| 48 | 12:10:22.266 | +16:47:12.45 | shallow | 6259 | 28.9 | ⋯ | ⋯ | 1150 | 26.5 | 1150 | 26.2 |
| 51 | 10:00:14.228 | +05:23:03.45 | deep | 15601 | 29.4 | 1500 | 26.9 | 2097 | 27.1 | 2385 | 26.8 |
| 53 | 10:03:46.445 | +32:53:48.35 | medium | 8340 | 29.2 | ⋯ | ⋯ | 1200 | 26.5 | 1300 | 26.2 |

**Note.** Fields that could not be included due to missing data are provided in Appendix A (Table 6). This table includes the PIE+ data (F475W for 18 fields as well as missing optical data for five fields) in addition to that from the original PIE program.

run UPDATEWCS on each exposure, which removes any attempted corrections to the astrometry applied by the calibration pipeline, and reverts to the astrometry provided by the HST guide stars. As these are based on the Gaia catalog (Gaia Collaboration et al. 2023), this provides the best baseline for our images and minimizes the difference in astrometry between each band.

We then combine our cleaned exposures into science images, also using the DRIZZLEPAC software (S. L. Hoffmann et al. 2021). This consists of the TWEAKREG routine, which is used to align the exposures on a consistent WCS grid, and ASTRODRIZZLE, which uses the reconstruction method known as "drizzling" (A. S. Fruchter & R. N. Hook 2002) to combine exposures into a final image. Note that, due to the lack of dithering in our fields, the point-spread function (PSF) is not subsampled, so we retain the native 40 mas pix$^{-1}$ scale in our images. Although the PIE+ data are dithered, accurate colors require pixel-matched images, so we also retain this pixel scale for PIE+ images throughout our processing.

We first consider the F814W exposures. We compare sources in each exposure with the Gaia catalog (Gaia Collaboration et al. 2023). TWEAKREG will then align the exposures to the coordinates of the Gaia sources, ensuring excellent alignment between the exposures. However, for a large fraction of our fields, there are fewer than three nonsaturated Gaia sources, so we cannot confidently align the images to the Gaia catalog. In these cases, we produce an initial "unaligned" drizzle. We use the source detection algorithm included with TWEAKREG to detect sources in this drizzle, and align each exposure with this drizzle. This means that, although absolute alignment is not confirmed using TWEAKREG, relative alignment is good to within ≈0.2 pixel or 8 mas. Once the individual F814W exposures are aligned, we use ASTRODRIZZLE to produce the final F814W science image for each field.

For the other filters, we produce an "unaligned" drizzle (although this is still quite accurate, as it use the Gaia-based guide star catalog), and then use TWEAKREG to adjust the unaligned drizzle to match the WCS of the F814W science image. This gives the required "tweak" for each individual exposure, which is then applied using the TWEAKBACK routine. We then use ASTRODRIZZLE as above to produce the final science image, but we drizzle the exposures onto the same pixel grid as the F814W, ensuring that both the WCS and the pixels of the filter-level drizzles are aligned.

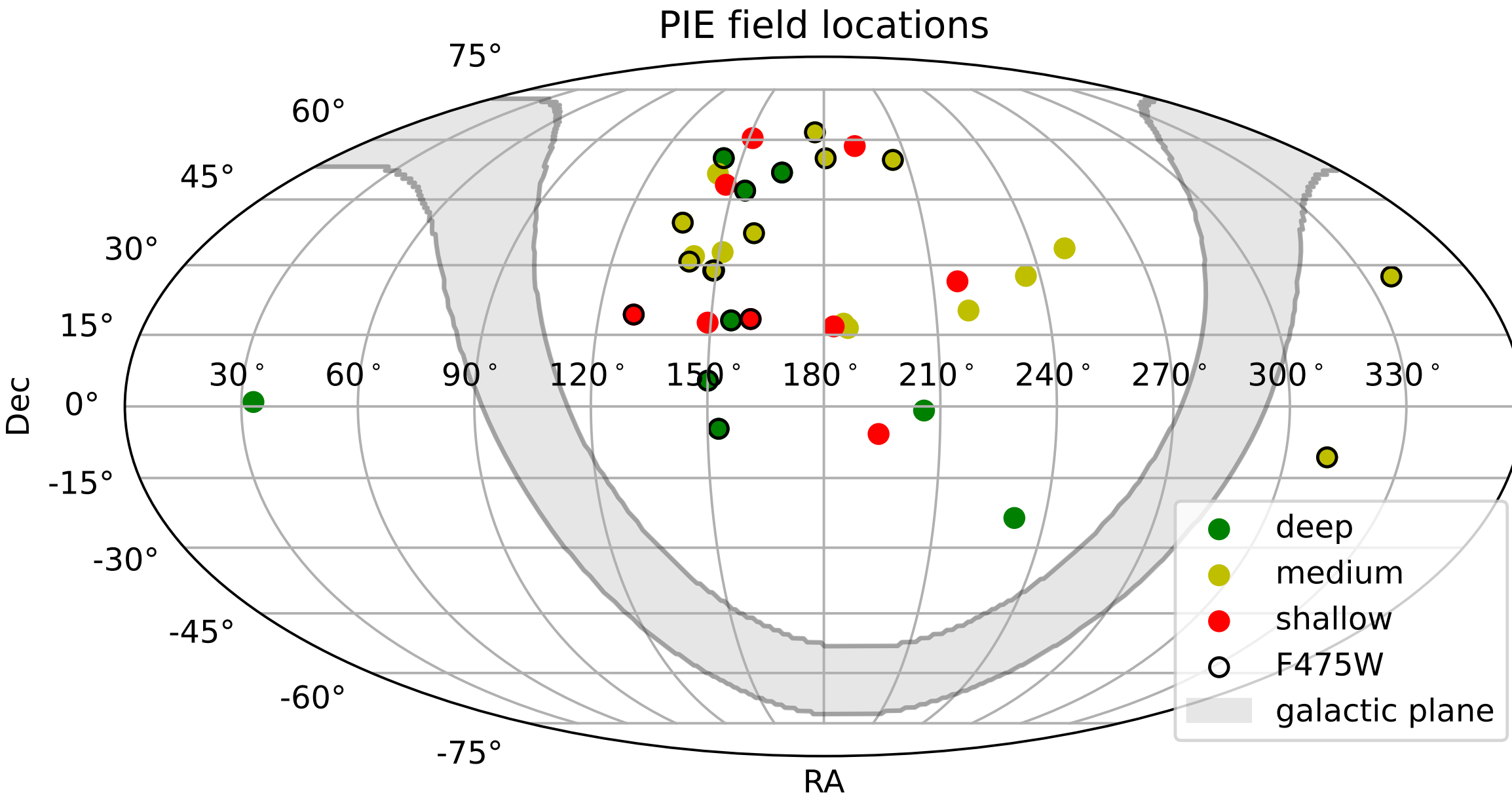

**Figure 2.** The locations of the PIE fields on the sky. Fields are divided into shallow (<2 hr), medium (2–4 hr), and deep (>4 hr), based on the usable F336W exposure time. Fields for which the PIE+ program contributed F475W observations are also circled in black. Regions within 10° of the galactic plane are shaded, as the extinction is generally too high in these regions for our UV observations to be useful.

With many different fields of different depths, the parameters used by TWEAKREG must also differ. These are varied by hand until no offset between the filters is visible (alignment better than ≈0.5 pixel).

Due to the lack of dithering in the PIE data, hot pixels in the detector lie at the same coordinates for all exposures, so these locations have no usable data. We flag these pixels and interpolate over them using the neighboring pixels so that they do not substantially affect the measured fluxes of our detected sources.

## 4. Simulations

The PIE program seeks to accurately recover LyC flux from LBGs at $3.1 < z < 3.5$ in the HST data discussed above. Here we compile and discuss three independent simulations we utilize to measure the robustness of our HST pipeline, color selection, and IGM attenuation assumptions in advance of the analysis of these data.

### 4.1. Jaguar

First, we use mock data from the JWST extragalactic mock catalog (JAGUAR; C. C. Williams et al. 2018) to motivate our color selection. The JAGUAR catalogs include mock spectra for over 300,000 galaxies, produced assuming observed galaxy mass and luminosity functions (for galaxy stellar masses $>10^6 M_\odot$ and redshifts between $z = 0.2$ and $z = 15$), which can be integrated through any HST filter in order to measure galaxy colors. Briefly, this uses BEAGLE (J. Chevallard & S. Charlot 2016) to generate mock spectral energy distributions (SEDs) using G. Bruzual & S. Charlot (2003) simple stellar populations. The emission from the H II regions is modeled using CLOUDY (G. J. Ferland et al. 2017), but these regions are assumed to be ionization-bounded and, therefore, allow no LyC escape. For this reason, mock galaxies in our target redshift range are all undetected in F336W. We discuss different methods for addressing this in Section 6.2. These model SEDs are attenuated by dust in the interstellar medium and by the intervening IGM, using prescriptions from S. Charlot & S. M. Fall (2000) and A. K. Inoue et al. (2014), respectively.

This provides a large catalog of ≈9000 mock galaxies with $m(\mathrm{F625W}) < 26.0$ (so are likely to be detected in our observations; see the depths listed in Table 1), which we can use to optimize our color selection (as described in Section 6).

### 4.2. PIE Mocks

We also create our own suite of mock galaxies. In addition to providing a second measurement of the completeness and purity of our color selection, we can inject these sources into our images and confirm that we can accurately recover the photometry using our pipeline. We select three fields into which we inject these mock sources. These are fields 22, 23, and 51, and are chosen to ensure that our "shallow," "medium," and "deep" fields are all represented, as well as at least one field with F475W observations.[13]

Our mock sources are created following commonly adopted methods (e.g., P. A. Oesch et al. 2010; S. L. Finkelstein et al. 2015; V. Mehta et al. 2017). Each mock galaxy is assigned a set of model parameters, including age, star formation history $\tau$, and dust extinction ($A_V$). The distributions for these model parameters are randomly generated from those observed in the 3D-HST survey (R. E. Skelton et al. 2014; K. E. Whitaker et al. 2014). These parameters are then assigned into Flexible Stellar Population Synthesis (FSPS; C. Conroy et al. 2009; C. Conroy & J. E. Gunn 2010) to generate a set of galaxy SEDs. FSPS includes both nebular emission and continuum. We include dust attenuation

[13] We note that these mocks were generated before the PIE+ observations were complete. The F475W data for field 51 was not available at that time and, therefore, could not be included.

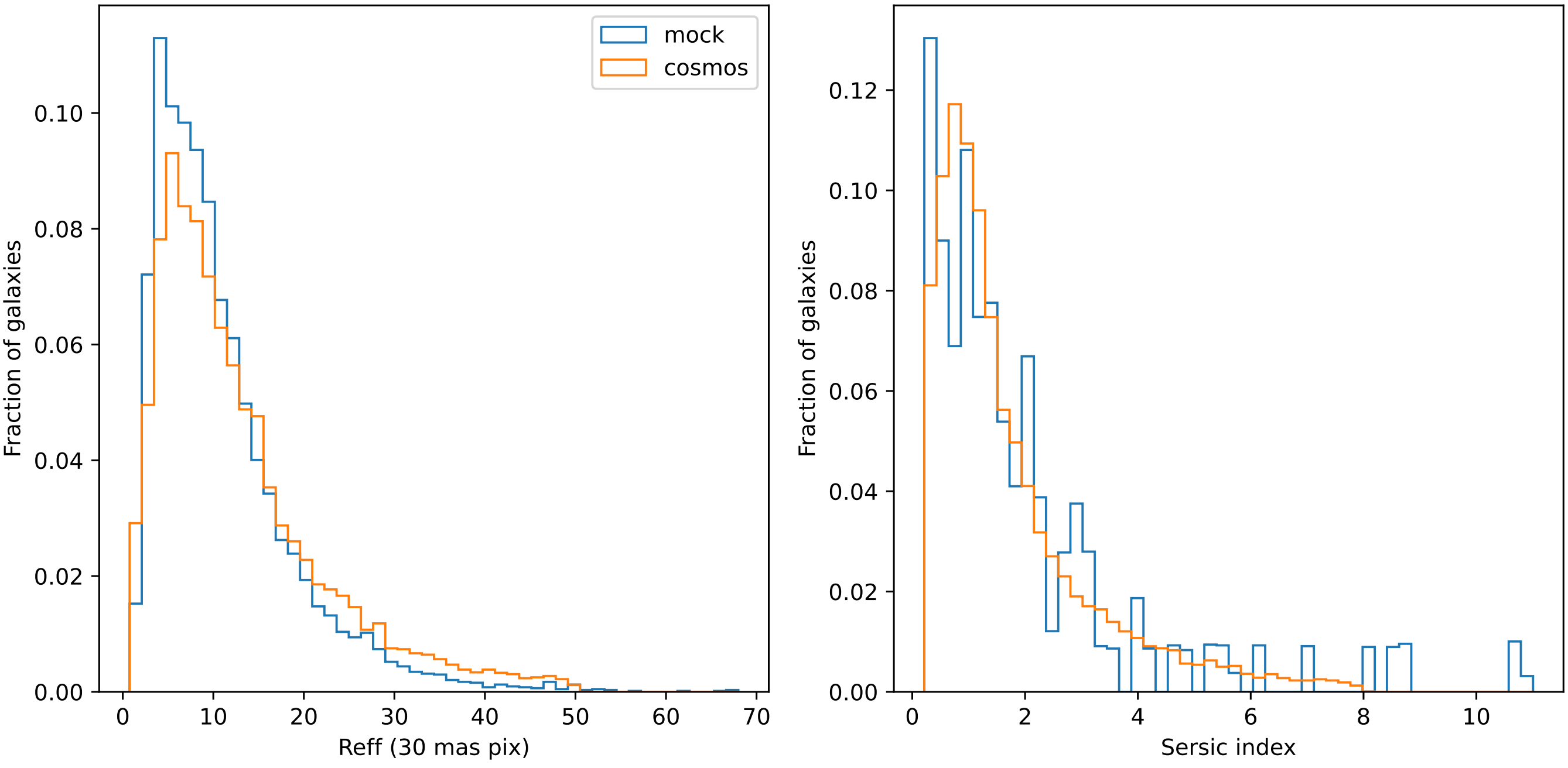

**Figure 3.** Comparison of the mock galaxy sizes (left panel) and Sérsic indices (right panel) with catalogs measured from the COSMOS field. Note that the PIE images have a different pixel scale than the COSMOS images, so here our mock sizes are scaled to match.

using the D. Calzetti et al. (2000) law, and IGM transmission using the curve from P. Madau (1995).

To replicate the observed galaxy number density, we assign the resulting spectra to a redshift range of 0.1–5.5 and then redshift the SEDs accordingly. We use the UV or *B*-band optical luminosity functions from C. Scarlata et al. (2007), V. Mehta et al. (2017), and R. J. Bouwens et al. (2015) for the redshift intervals $0.1 < z < 1.5$, $1.5 < z < 3.5$, and $3.5 < z < 5.5$, respectively. Only galaxies brighter than 26.5 mag in F625W are simulated.

Galaxy sizes are assigned using the observed mass–size relation from K. V. Nedkova et al. (2024), with Sérsic indices drawn from the UVCANDELS catalog described in V. Mehta et al. (2024). We do not find a significant correlation between galaxy sizes and Sérsic indices in UVCANDELS, so the values of these two parameters used in our mocks are drawn independently from the UVCANDELS distribution.

In Figure 3, we show the size and Sérsic index distributions from the COSMOS catalog and from the mock catalog generated using this method. The mock and observed distributions are very similar for both parameters, although the distribution of mock Sérsic indices is clearly more stochastic. There are some minor, but visible, differences in the size distribution. At small sizes, this may be due to the difficulty of measuring accurate sizes for unresolved sources in the COSMOS field.

We create 100 realizations for each of the selected fields. Each realization contains 120 mock sources with magnitudes determined from the mock spectra, injected into all HST bands from that field, leading to a total of 12,000 mock sources. These are convolved with the PSF measured for that image using the procedure described in Section 5, in order to produce realistic sources from which we can validate our photometric measurements and measure completeness.

### 4.3. Mock IGM

In addition to our two sets of mock galaxies, we utilize simulations of the IGM in order to estimate the fraction of LyC flux emitted by galaxies in our sample that is absorbed by the intervening IGM. We estimate the probability of a given opacity using the method described in R. Bassett et al. (2021).

To summarize, this method populates mock sight lines with H I absorbers using the redshift-dependent column density distribution function from C. C. Steidel et al. (2018), covering column densities between $10^{12}$ and $10^{21}$ cm$^{-2}$. Redward of the Lyman limit, Voigt absorption profiles are fit to these absorbers, whereas at bluer wavelengths, the absorption is proportional to $N_{\rm HI}\lambda^3$ (e.g., D. E. Osterbrock 1974). This gives the full UV absorption spectrum of the sight line as a function of wavelength. We generate 10,000 mock sight lines at $z = 3.1$ and 10,000 at $z = 3.5$.

In the left panel of Figure 4, we show the median IGM transmission ($T_{\rm IGM}$) as a function of observed wavelength at the limits of our target redshift range. We also indicate the 10th and 90th percentiles, as well as the F336W filter through which we aim to detect LyC emission. Not only is there a wide range of $T_{\rm IGM}$ values at both redshifts but a substantial change in $T_{\rm IGM}$ across the relatively small target redshift window. This means that a clear line of sight to a galaxy at $z = 3.1$ is far more likely than to a galaxy at $z = 3.5$, in the wavelengths visible through F336W, due to the rapid decrease in IGM transmission blueward of the Lyman Limit. We show the IGM transmission in F336W in the right panels, which demonstrate that very few sight lines allow 50% of F336W flux through the IGM. At both redshifts, a substantial fraction of sight lines are completely opaque, so we will not be able to detect LyC emission in these sight lines. However, many sight lines are still somewhat transparent at both redshifts, with around 20% of sight lines at $z = 3.5$ and 50% of sight lines at $z = 3.1$ having IGM transmission >10%. The median IGM transmission is found to be ≈7% at $z = 3.1$ and ≈0.3% at $z = 3.5$. With the large number of independent fields in the PIE survey, we can assume these values of average IGM transmission without biasing our results.

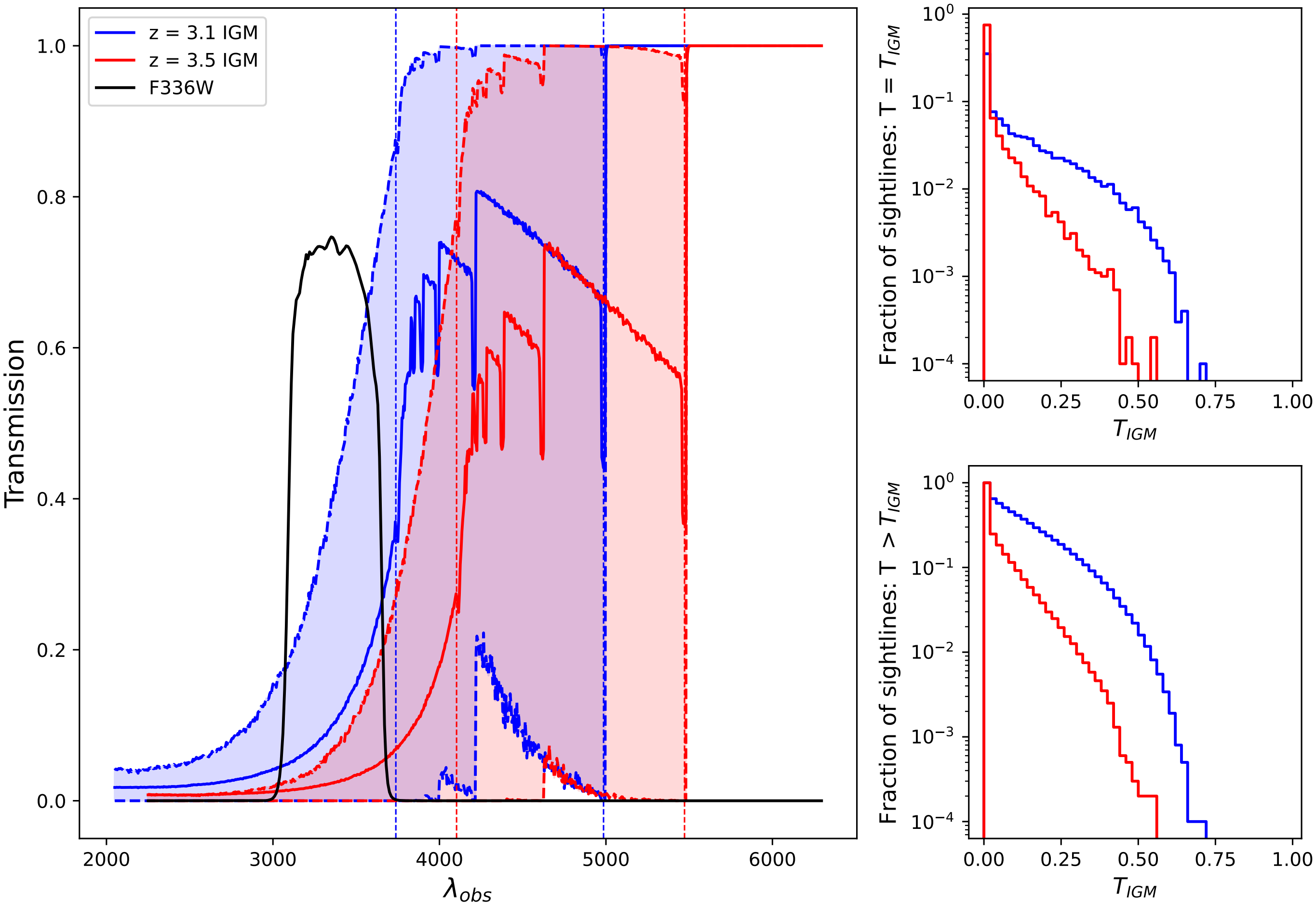

**Figure 4.** Expected transmission of the IGM at $z = 3.1$ and $z = 3.5$, using the method from R. Bassett et al. (2021). Left panel: median IGM transmission as a function of observed wavelength (solid line). The dashed lines outlining the shaded regions show the 10th and 90th percentiles, and the F336W filter throughput is also shown in black. Vertical dashed lines also indicate the Lyman limit and the wavelength of Ly$\alpha$ for both redshifts. Top-right panel: histogram of average IGM transmission through the F336W filter for each of the 10,000 simulated sight lines, showing the fraction of mock sight lines in each bin of $T_{\mathrm{IGM}}$. Bottom-right panel: cumulative histogram of $T_{\mathrm{IGM}}$, showing the fraction of sight lines more transparent than the given value.

## 5. Photometry

Accurate photometry is vital for selecting the maximum number of LBGs for our spectroscopic follow-up and, hence, maximizing the size of our final sample. Here we describe the measurement and validation of our galaxy photometry.

### 5.1. Measurement

Galaxies were identified in our images using the SExtractor software (E. Bertin & S. Arnouts 1996). SExtractor is a flexible tool for detecting and deblending astronomical sources, as well as measuring their photometry. It does this by detecting adjacent pixels above a defined threshold, uses their fluxes to deblend them into individual sources, and assigns each pixel to a single source. The resulting segmentation map can be used to measure the fluxes of those same sources in all of our bands. Ensuring accurate colors requires PSF-matching the images to equalize the fraction of flux lost outside of the measured regions.

In order to model the PSF, we add 250 point sources at a range of subpixel locations into the exposure-level images. These point sources are generated using the WFC3 library PSF for the relevant filter at the focus value corresponding to the time when the exposure was taken. These model exposures are then drizzled to produce filter-level images with the same pixel size and orientation as the real drizzled images. We then extract the model point sources, interpolate onto a finer grid with pixels 20 times smaller than the images (2 mas), and fit them with a Gaussian model. We then align the centroids of these models, take the median value in each smaller pixel, and re-scale them back onto the correctly sized pixel grid. This process ensures that we have enough sources to subsample the PSF, which is not guaranteed if we use real, nonsaturated stars in the images.

We use Photutils on these model PSFs to create convolution kernels for each filter and each F336W/F475W/F625W image, which is then convolved with that image to produce an image PSF-matched to F814W, ensuring that our measured colors are not affected by losses due to the PSF.

Our photometry is measured using SExtractor in dual-image mode, with a white-light detection image produced by coadding the PSF-matched F625W and F814W images. As in other studies of distant galaxies (e.g., M. Rafelski et al. 2015; M. Revalski et al. 2023), we require a minimum area of 9 contiguous pixels, approximately corresponding to the 3 x 3 pixels of the FWHM of the PSF. The detection threshold (DETECT THRESH) is set to capture as many sources as possible while excluding spurious sources due to noise. We determine this by inverting the detection image (multiplying every value by −1), masking out any apparent crosstalk artifacts,

**Table 2**
SExtractor Parameters Used for Photometry of the PIE Fields

| Parameter | Value | Unit |
|---|---|---|
| DETECT MINAREA | 9 | pixels |
| DEBLEND NTHRESH | 32 | … |
| DEBLEND MINCONT | 0.005 | … |
| DETECT THRESH | ([a]) | … |
| ANALYSIS THRESH | ([a]) | … |
| BACK SIZE | 64 | pixels |
| BACK FILTERSIZE | 3 | pixels |
| CLEAN PARAM | 1.0 | … |
| PHOT AUTOPARAMS | 2.5, 3.5 | $R_{\mathrm{kron}}$, pixels |

**Note.** The determination of other parameters is discussed in the text.
[a] DETECT THRESH and ANALYSIS THRESH are set by an iterative process described in the text.

and running SExtractor on the negative image, as all "sources" in the negative image must be artificial. Beginning from a $0.5\sigma$ threshold, we attempt to detect sources in the negative image, increasing this threshold in steps of 0.25 until we find the lowest value at which no sources are found (usually DETECT THRESH $\approx 2$), and therefore the lowest threshold that will eliminate spurious sources when run on the science images. The analysis threshold (ANALYSIS THRESH) is set to the same level as the detection threshold. We discuss the completeness of the resulting catalog in Section 5.3.

The photometric zero-point (MAG ZEROPOINT) and pixel scale (PIXEL SCALE) are determined from the image header, the GAIN is set to the total exposure time for that image, and the PSF width is measured from the model PSF for that filter (SEEING FWHM). The remaining SExtractor parameters are kept to the default values,[14] which are already optimized for source detection. For reference, some of the relevant parameters are provided in Table 2.

### *5.2. Validation*

We validate our photometry using the mock galaxy sample described in Section 4.2. By comparing the magnitudes of the sources that were implanted into the HST images with those extracted using our photometry pipeline, we can confirm that sources are being extracted correctly.

By default, SExtractor uses an aperture set at 2.5 times the Kron radius (R. G. Kron 1980). This theoretically encloses >90% of the total light from a galaxy. However, this assumes that the light profile is integrated to a sufficient distance to obtain an accurate Kron radius. In measured sources, the light profile is not integrated to this extent because the outer edges of the galaxy do not significantly exceed the noise floor. Hence, the Kron radius is underestimated, and the recovered flux is lower, as has been found in several previous studies (e.g., N. Benítez et al. 2004; A. W. Graham & S. P. Driver 2005; D. Hammer et al. 2010). This effect is more significant for sources with a high Sérsic index, which have more flux at large radii that could be missed from the Kron aperture. Thus, while we would expect an aperture correction of ≈0.1 mag, the actual aperture correction is likely to be higher. This correction can be measured using our mock galaxies.

[14] We use SExtractor version 2.12.4.

We match our SExtractor results with the mock catalogs for sources within 3 pixels and compare the Kron magnitudes produced by SExtractor with the total input magnitudes. As we are targeting high-redshift sources, we expect our target galaxies to be small and compact; thus, we optimize our correction for these galaxies. We therefore only consider galaxies that are fairly compact (effective radius < 8 pixels, ≈2.5 kpc at $z \approx 3$), with a low Sérsic index ($n < 1.8$). These cuts include most galaxies at our target redshift while removing low-$z$ extended sources (e.g., K. Ormerod et al. 2024; E. Ward et al. 2024). Using sources with the same size as our targets will allow the correction to also capture any effects caused by the PSF. In order to avoid our correction being distorted by faint sources that are not well measured, we also restrict this comparison to sources with high signal-to-noise (above 40, so around 24th magnitude and brighter).

The results for field 51 (the medium-depth field) are shown in Figure 5, and the results for all three mock fields are presented in Table 3. In each case, our measured magnitudes are ≈0.2 mag fainter than the total input magnitudes, although the recovered colors are accurate. Using a lower-S/N cut returned similar values for the magnitude offsets, although with a larger spread, showing that the correction is not strongly magnitude-dependent.

This ≈0.2 mag offset required to align our measured magnitudes with the simulations is, as expected, larger than the theoretical 0.1 mag correction between Kron magnitudes and total magnitudes. However, this offset is consistent between the different bands, so our measured colors are found to be accurate to within 0.03 mag in field 51, 0.04 mag in field 22, and 0.05 mag in field 23. Although the uncertainty increases when lower signal-to-noise sources are included in the comparison, the median difference between the mocks and measurements does not shift significantly. The measured uncertainties are consistent with the noise levels at these signal-to-noise values, suggesting no correction or additional uncertainty is required on our colors.

As the magnitude offsets shown in the left panel of Figure 5 demonstrate, the difference between the offsets in the different bands is smaller than the scatter in this offset. Thus, combining these values into a single aperture correction applied to the PSF-matched images allows us to recover a smaller uncertainty than correcting each filter separately.

We note that it is common for the isophotal magnitudes measured by SExtractor to be used in measuring galaxy colors (N. Benítez et al. 2004; D. Coe et al. 2013; M. Rafelski et al. 2015; H. Yan et al. 2023). The isophotal magnitudes are usually measured from a smaller aperture than the Kron-based "AUTO" magnitudes, so generally have a higher signal-to-noise level but are expected to require a larger aperture correction in order to recover the total flux. However, the differences between the measured isophotal and mock magnitudes, as well as being larger, are also less consistent between the different fields and filters, and have a larger scatter than the "AUTO" magnitudes. This leads to color offsets of ∼0.2 mag in "ISO" magnitudes, compared with the <0.05 mag seen using the "AUTO" magnitudes. Therefore, we use the Kron-based apertures in our catalogs and analysis.

### *5.3. Detection Completeness*

It is also important to determine how deep our measurements extend, so we utilize the PIE mocks described in Section 4.2 to characterize our completeness. As discussed in Section 5.1, we

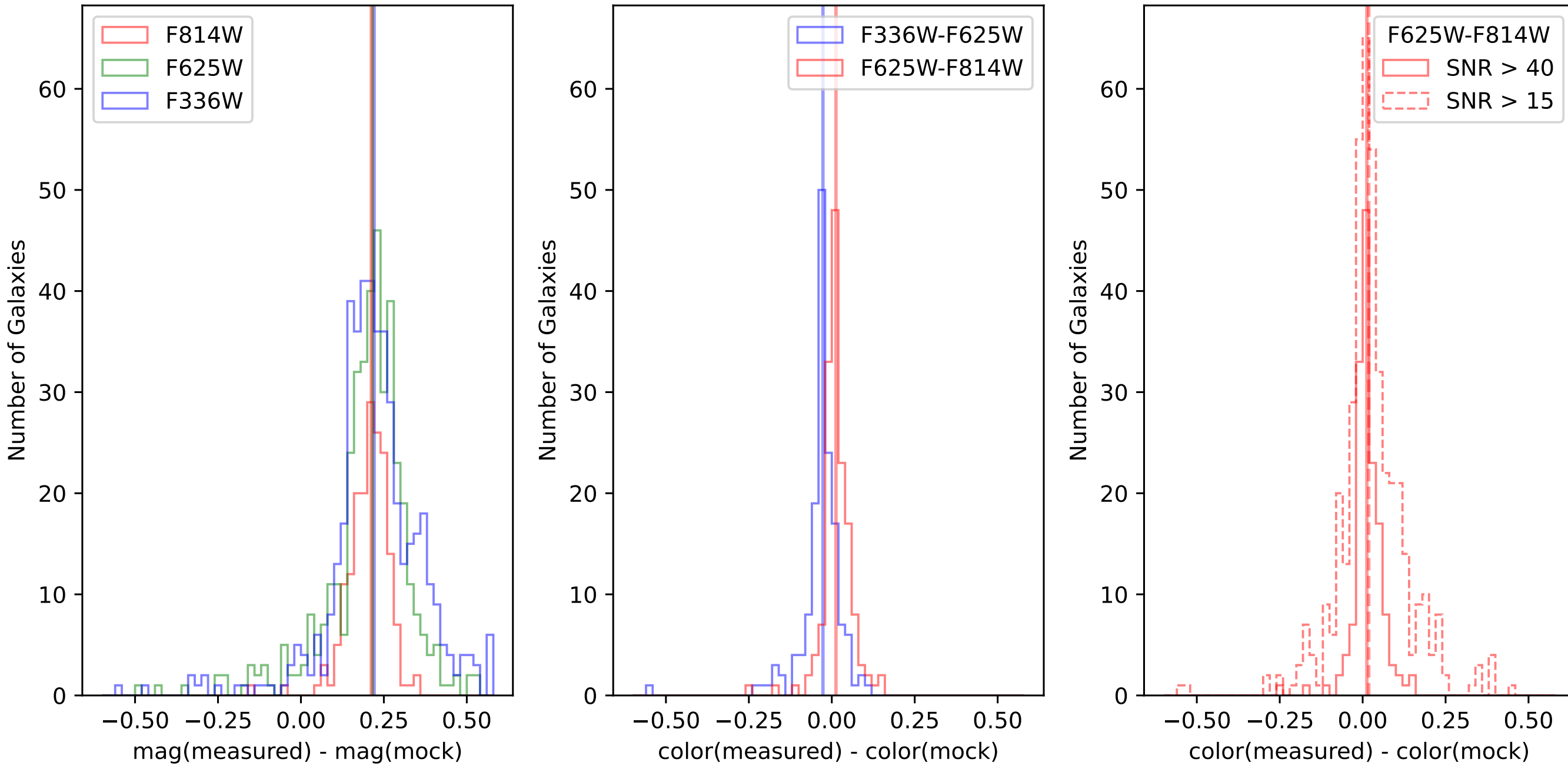

**Figure 5.** Color and magnitude offsets between the mock input catalog and the SExtractor measurements on the mock images, with sources curated as described in Section 5.2. Field 51 (medium-depth) is shown, but results are similar for the other mock fields. Left panel: magnitude offsets in the three PIE filters. Middle panel: color offsets in the two colors used for our primary selection of high-redshift galaxies. Right panel: F625W − F814W color offsets using sources above different signal-to-noise thresholds. The median of each distribution is shown by a vertical line. The median and spread of the distribution are given in Table 3.

**Table 3**
Color and Magnitude Differences between the Catalog of Simulated Galaxies and Measured Sources

| Property | Field 22 | | Field 23 | | Field 51 | |
|---|---|---|---|---|---|---|
| | Offset (mag) | $\sigma$ (mag) | Offset (mag) | $\sigma$ (mag) | Offset (mag) | $\sigma$ (mag) |
| F336W magnitude | 0.23 | 0.13 | 0.21 | 0.09 | 0.21 | 0.10 |
| F475W magnitude | 0.23 | 0.10 | … | … | … | … |
| F625W magnitude | 0.23 | 0.10 | 0.20 | 0.06 | 0.21 | 0.09 |
| F814W magnitude | 0.21 | 0.07 | 0.16 | 0.03 | 0.20 | 0.06 |
| F336W − F475W color | −0.01 | 0.01 | … | … | … | … |
| F336W − F625W color | −0.04 | 0.02 | −0.01 | 0.02 | −0.03 | 0.03 |
| F475W − F625W color | −0.02 | 0.02 | … | … | … | … |
| F625W − F814W color (S/N > 40) | 0.01 | 0.03 | 0.03 | 0.03 | 0.01 | 0.03 |
| F625W − F814W color (S/N > 15) | 0.03 | 0.08 | 0.05 | 0.07 | 0.02 | 0.08 |

**Note.** The mock images were analyzed using the same pipeline as discussed in the text. Note that $\sigma$ denotes the scatter in the distribution and, hence, the uncertainty that needs to be included on the measurement for any individual galaxy.

use a detection image formed by coadding the F625W and F814W PSF-matched images for each field. We then measure the fraction of mock sources that are detected as a function of magnitude, using the SExtractor-based pipeline discussed above. In order to ensure that our completeness is not affected by any difficulty centroiding very extended sources or the associated flux loss from the Kron-based apertures, we again cut the mock sample to only include sources with effective radius $< 8$ pixels and Sérsic index $n < 1.8$.

In Figure 6, we show the fraction of mock sources detected as a function of their magnitude (as given in the mock catalog) for each filter and field. Detected sources are those that meet the detection threshold from our white-light images (so are included in our photometric catalogs) and are detected with $S/N > 3$ in the relevant filter. We find that our fields with half-orbit depth in F625W and F814W give 50% completeness at $m$(F625W) $\approx$ 25.6, with deeper fields reaching $m$(F625W) $\approx$ 26.4. However, the depth of our F336W often allows $3\sigma$ detections down to $m \approx 28$.

We can also confirm the depth of our catalog by comparing the density of sources in our fields with other surveys. Figure 7 shows the density of galaxies in our catalog as a function of F625W magnitude, and compares with density in F606W from the UVUDF galaxy catalog (M. Rafelski et al. 2015) and a fit to multiple surveys (S. Koushan et al. 2021). These data are much deeper than PIE in the optical bands, so a smaller number of detected sources corresponds to an incompleteness.

The PIE source counts are consistent with (or slightly higher than[15]) the deeper data to $m$(F625W) $> 25.0$, and reach 50% completeness at about $m$(F625W) $= 26.0$, consistent with the results derived from the mock source extractions. The "tail" of faint sources is expected due to our use of a white-light detection image, with these sources bright enough in F814W to reach the detection threshold in the white-light image.

[15] Some of our fields feature overdensities, leading to a higher source count that the UVUDF.

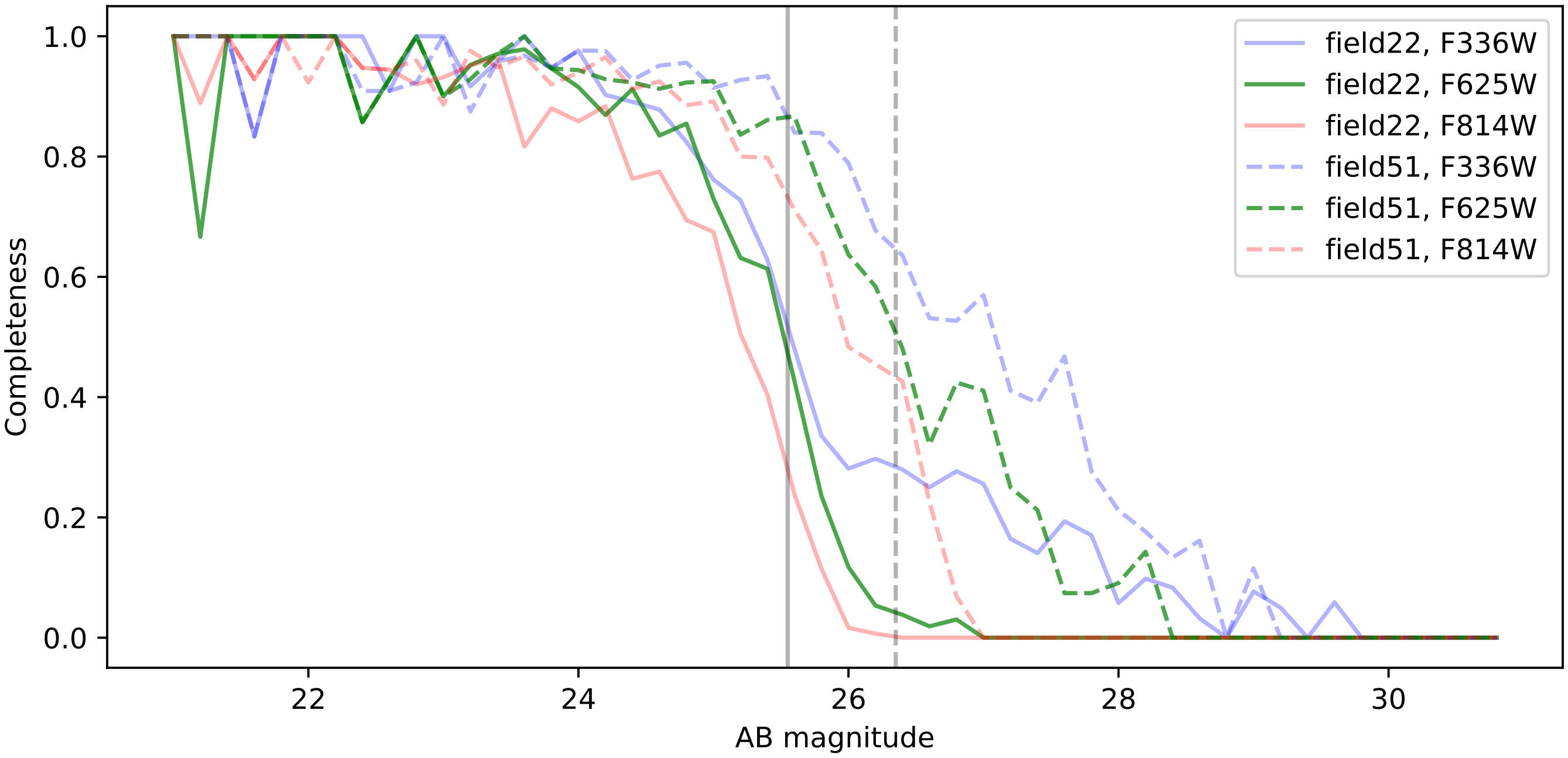

**Figure 6.** Detection completeness of our sample, estimated using the fraction of mock sources that are detected at the $3\sigma$ level when searching the implanted images. Vertical lines indicate the magnitude at which we achieve ≈50% completeness in F625W.

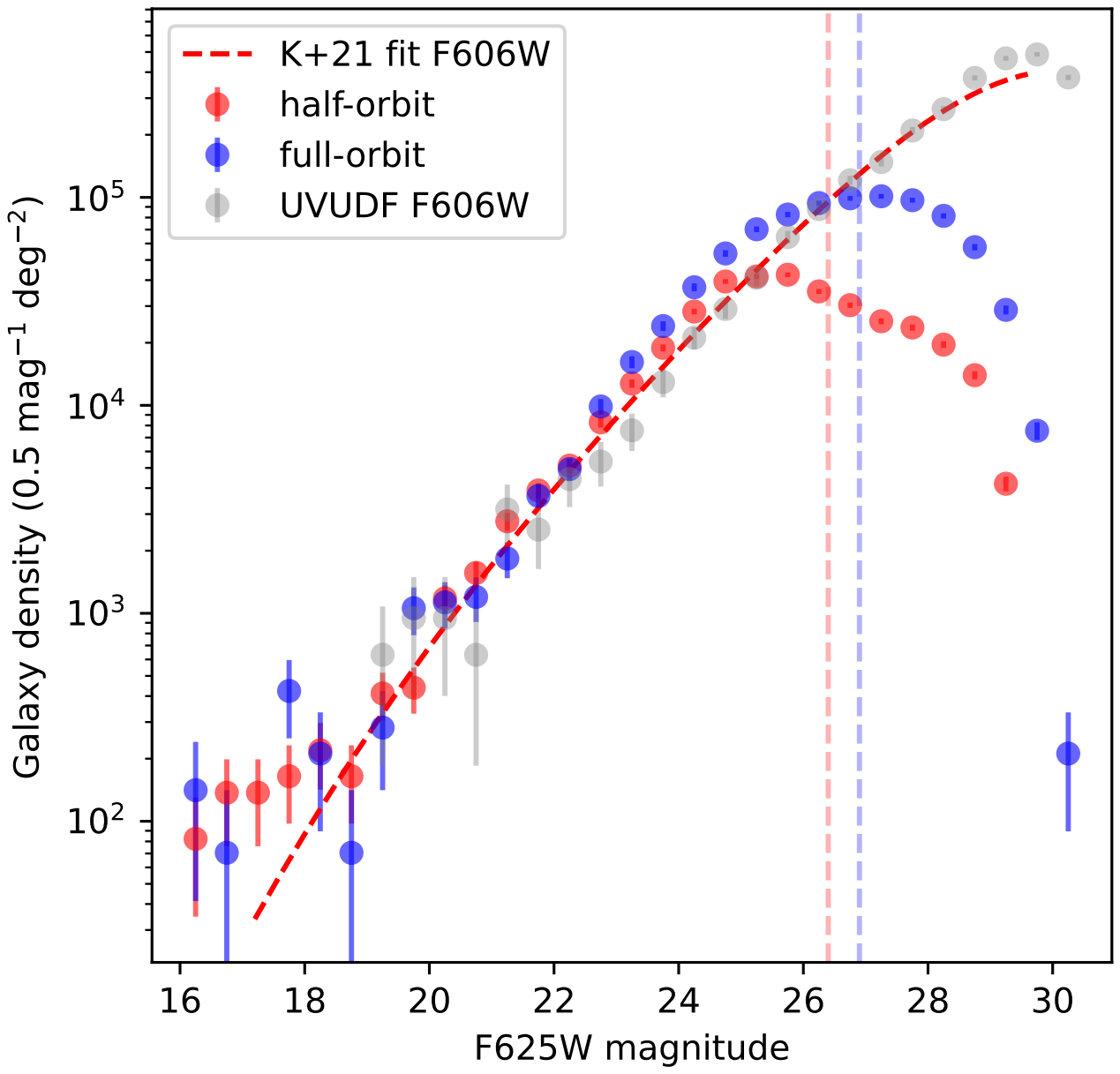

**Figure 7.** Average density of galaxies in the PIE fields. We show only fields observed in F814W and F625W with the PIE program (as PIE+ data is usually at a different orientation, affecting the overlapping area), split into fields with half-orbit and full-orbit depth in F625W. Also included are the densities of F606W sources from the UVUDF galaxy catalog presented in M. Rafelski et al. (2015) and the fit to several galaxy surveys given by S. Koushan et al. (2021). For comparison, the vertical dashed lines show the typical $5\sigma$ depth of half-orbit and full-orbit F625W exposures, as presented in Table 1.

## 6. Color Selection

### 6.1. Optimized LBG Selection

In advance of the spectroscopic campaign, we need to select a sample of target galaxies in order to maximize our sample size and minimize the number of low-redshift interlopers. We use both the JAGUAR and PIE mocks described in Sections 4.1 and 4.2 in order to optimize our color selection.

For both of these mock data sets, we first remove all galaxies fainter than $m$(F625W) = 26. In most cases, this ensures S/N > 7 in this band, so that we can reliably measure photometry for similar sources in our data. (As shown in Table 1, we reach $5\sigma$ depth of ≈26.5 in most fields.) For galaxies that are undetected in F336W, we assume a $1\sigma$ detection limit of 29.0 mag. In our HST imaging, this is obtained in about 8000 s of F336W exposure, which, as can be seen in Table 1, is reached in most of the PIE fields. Additionally, neither mock sample features any galaxies in our target redshift range with $m$(F625W) < 23.5, so we can remove bright sources from our target sample without any loss in completeness.

We then use a color selection based on F336W − F625W and F625W − F814W colors, in order to identify LBGs at $z \gtrsim 2.7$. Similar cuts have long been used to identify LBGs (e.g., C. C. Steidel et al. 2003; M. Rafelski et al. 2009; N. P. Hathi et al. 2010; P. A. Oesch et al. 2013; H. I. Teplitz et al. 2013). We optimize our selection to maximize the completeness (high-redshift galaxies selected/total high-redshift galaxies) and purity (high-redshift galaxies selected/all galaxies selected) and, hence, select the maximum number of target galaxies for spectroscopic follow-up.

The results are shown in the upper panels of Figure 8, with the completeness and purity of each selected sample listed in Table 4. The color cuts shown (outlined in green) are:

$$
\begin{aligned}
-0.45 < (\mathrm{F625W} - \mathrm{F814W}) &< 0.4 \\
(\mathrm{F336W} - \mathrm{F625W}) &> 1.5 \\
(\mathrm{F336W} - \mathrm{F625W}) > 2.4\,(\mathrm{F625W} - \mathrm{F814W}) &+ 1.9.
\end{aligned}
$$

This selection captures ≳95% of galaxies at $2.7 < z < 4.0$ (red circles selected, red crosses missed), with a purity of ≈80% (interlopers outside this redshift range shown as blue crosses).

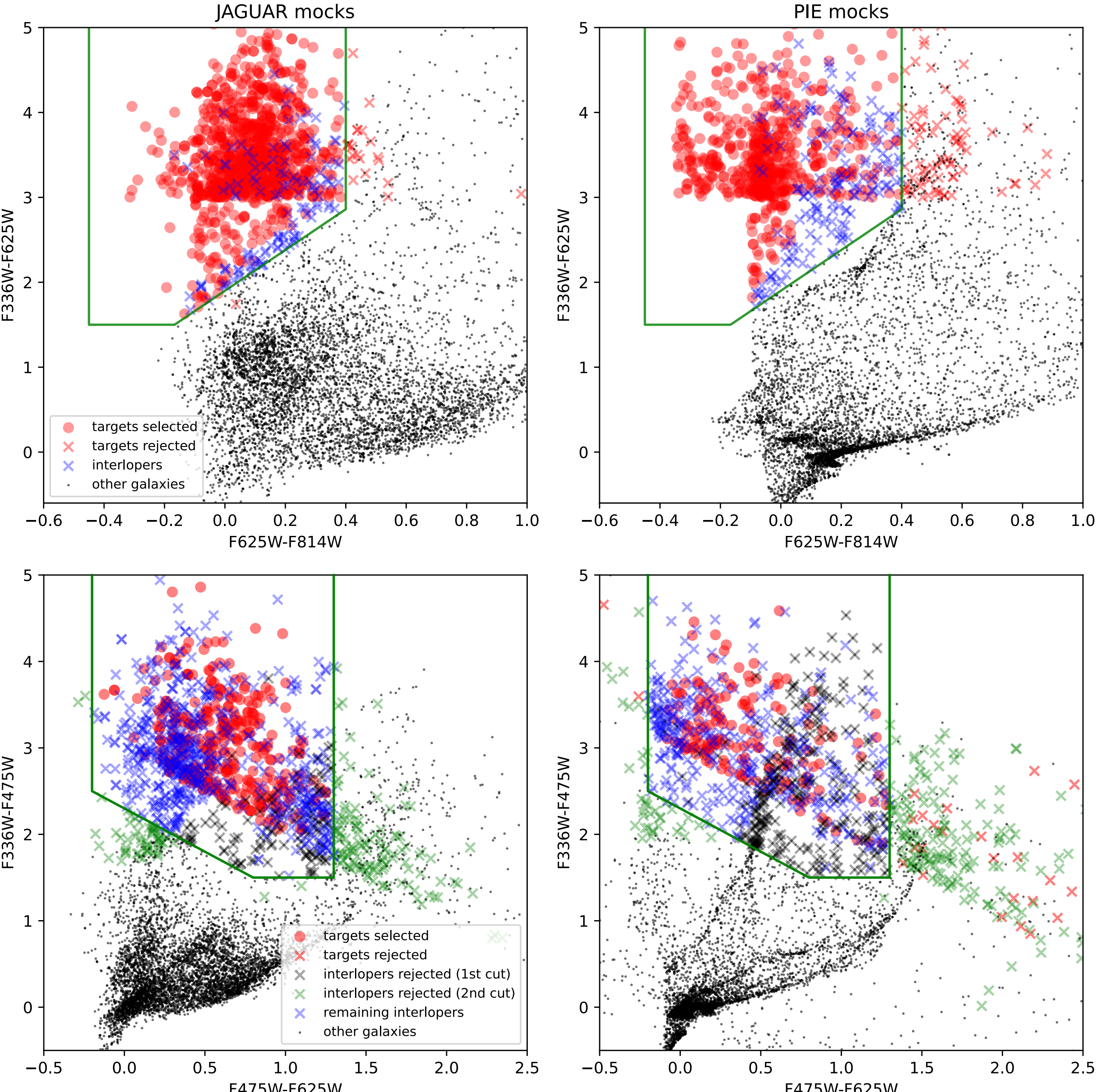

**Figure 8.** Color selection using the JAGUAR and PIE mock data described in Section 4. Top panels: color selection based on F336W, F625W, and F814W data, which will be used for fields for which only these filters are available. Bottom panels: additional selection criteria that will be used to find $3.1 < z < 3.5$ galaxies for fields with F475W PIE+ data in addition to the three PIE filters. In the top panels, red points are galaxies at $2.7 < z < 4$, whereas in the bottom panels, these are galaxies at $3.1 < z < 3.5$. In both cases, blue crosses are galaxies that would be selected using the color cuts shown by the green lines and given in the text, but are outside this redshift range. Galaxies that would be selected using our three-filter cuts but are removed by the additional cuts are shown in green in the lower panels. Other galaxies, lying outside the target redshift range and outside our selection window, are shown in black points, with crosses denoting those that lie inside the PIE + cuts but were already rejected using the PIE 3-band cuts.

We note that Figure 8 shows a large concentration of sources at (F336W − F625W) $\approx$ 3–4. This is due to the faintest galaxies in the optical (F625W mag $\approx$ 26) often going undetected in F336W. Any galaxies with detectable LyC emission will have a bluer F336W − F625W color, so more stringent cuts in this area are not desirable, even though this would result in a higher purity.

Similar studies often do not select against sources with a very blue optical color (F625W − F814W in our case). N. P. Hathi et al. (2010) used such a cut to account for the large difference in depth between their bands probing the UV continuum. We include this cut as well, even though there is a much smaller difference in depth between our bands, since no galaxies with these blue optical colors are present in either of our mock samples. This implies that any such sources observed are likely to be spurious. We use a slightly bluer cut than suggested by the mocks to allow for some uncertainty in our photometric measurements.

However, we note that selection criteria based on two colors can only select a very broad redshift range, so only about one-third of the galaxies selected lie at the $3.1 < z < 3.5$ target redshift range for which we can use F336W to measure LyC emission. Figure 9 shows this more clearly, with the completeness of our selection in both mock data sets very high between $2.7 < z < 4.0$ (yellow vertical band). Very few galaxies at lower redshifts are selected, so our color cuts are very efficiently removing low-redshift galaxies from our sample.

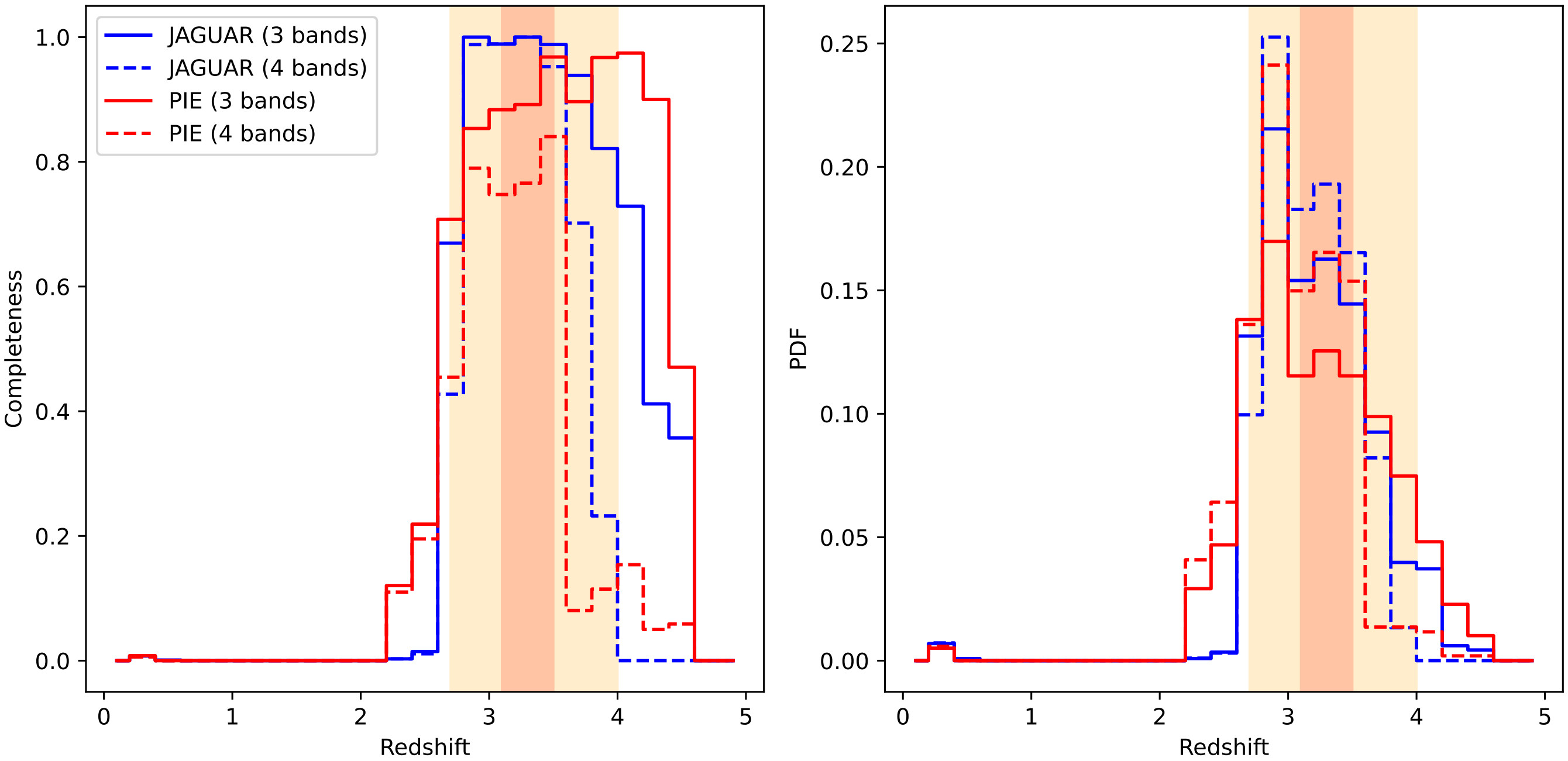

**Figure 9.** Redshift dependence of galaxies selected by the color cuts shown in Figure 8. These are shown in redshift bins of $\Delta z = 0.2$. Left panel: fraction of galaxies in each redshift bin that are selected by the color cuts. Right panel: fraction of galaxies selected by the color cuts that lie in each redshift bin. Shaded regions in both panels represent the $2.7 < z < 4.0$ range for which we optimize the three-band cuts (yellow), and the $3.1 < z < 3.5$ target redshift range which we aim for with the four-band cuts (red). Solid lines show the results using the three-band color selection, and dashed lines show results when the F475W filter is also used.

For the 18 fields with PIE+ observations in F475W, an additional color criterion can be used to significantly narrow the redshift range of the selected sources, and thereby increase the fraction of sources for which we can measure LyC emission. In the lower panels of Figure 8, we show the color cuts chosen using F475W to optimize for $3.1 < z < 3.5$ galaxies. These are:

$$-0.1 < (\mathrm{F475W} - \mathrm{F625W}) < 1.2$$
$$(\mathrm{F336W} - \mathrm{F475W}) > 1.5$$
$$(\mathrm{F336W} - \mathrm{F475W}) > -0.8(\mathrm{F475W} - \mathrm{F625W}) + 2.3.$$

By removing some galaxies at $2.5 \lesssim z \lesssim 3$ and at $\gtrsim$3.5, the purity of our target sample is increased from ≈30% to ≈40% for these fields. This is shown by the dashed lines in Figure 9, and allows us to target more of these sources with follow-up spectroscopy.

### *6.2. LyC Emission*

As we note above, the mocks used to optimize our color selection do not include any LyC emission, instead assuming $f_{\mathrm{esc}} = 0$. Strong LyC leakers could therefore shift out of our selection window, causing us to miss them from our spectroscopic sample.

To address this, we simulate the effect of adding LyC flux to our mocks. Following other studies of LyC emission (e.g., J. Cooke et al. 2014; L. J. Prichard et al. 2022), we define $R_{\mathrm{obs}} = (F_{\mathrm{LyC}}/F_{\mathrm{UV}})_{\mathrm{obs}}$. This is the observed ratio between flux density blueward of the Lyman limit and flux in the UV continuum (near 1500 Å), and is by definition the product of the intrinsic decrement between LyC and UV emission, the escape fraction $f_{\mathrm{esc}}$, and the transmission of the IGM. This

**Table 4**
Completeness and Purity Obtained when Selecting Galaxies from Several "Mock" Data Sets Using the Selection Described in Section 6 and Shown in Figure 8

| Data Set | Selected | Completeness | Purity |
| --- | --- | --- | --- |
| $2.7 < z < 4.0$ | (3 bands) | ... | ... |
| JAGUAR | 1156 | 0.98 | 0.89 |
| PIE mocks | 969 | 0.89 | 0.78 |
| $3.1 < z < 3.5$ | (4 bands) | ... | ... |
| JAGUAR | 903 | 0.99 | 0.38 |
| PIE mocks | 514 | 0.82 | 0.33 |

**Note.** Completeness is defined as the fraction of galaxies in the specified redshift range that are selected. Purity is the fraction of selected galaxies that are in the stated redshift range.

corresponds to $R_{\mathrm{obs}} = (F_{\mathrm{F336W}}/F_{\mathrm{F625W}})$ for our target galaxies at $3.1 < z < 3.5$.

We assume a constant level of LyC flux corresponding to values of $R_{\mathrm{obs}}$ from 0.01–0.2, which sets the F336W magnitude where this exceeds the $1\sigma$ limit shown previously. We then re-measure the colors and plot their location in color space in Figure 10.

The two panels of this figure show that adding LyC flux at levels below $R_{\mathrm{obs}} = 0.1$ does not cause a large fraction of galaxies to shift out of our selection window (we measure a completeness of ≈95% with the three-band or the four-band cuts), although most galaxies would shift out of our selection at $R_{\mathrm{obs}} = 0.2$ (completeness drops to <10%).

The intrinsic LyC/UV luminosity ratio depends heavily on the age of a galaxy (J. Chisholm et al. 2019), but is usually smaller than 1 for stellar ages $\gtrsim$10 Myr (and, hence, will be smaller than 1 for the large majority of our galaxies). As we show in Section 4.3, typical IGM transmissions are smaller than 20% for most sight lines, with median values of 7% and 0.3% at $z = 3.1$ and 3.5, respectively. Hence, even if $f_{\rm esc} = 1$, most of our galaxies will still have $R_{\rm obs} < 0.1$, so will not be shifted out of our color selection.

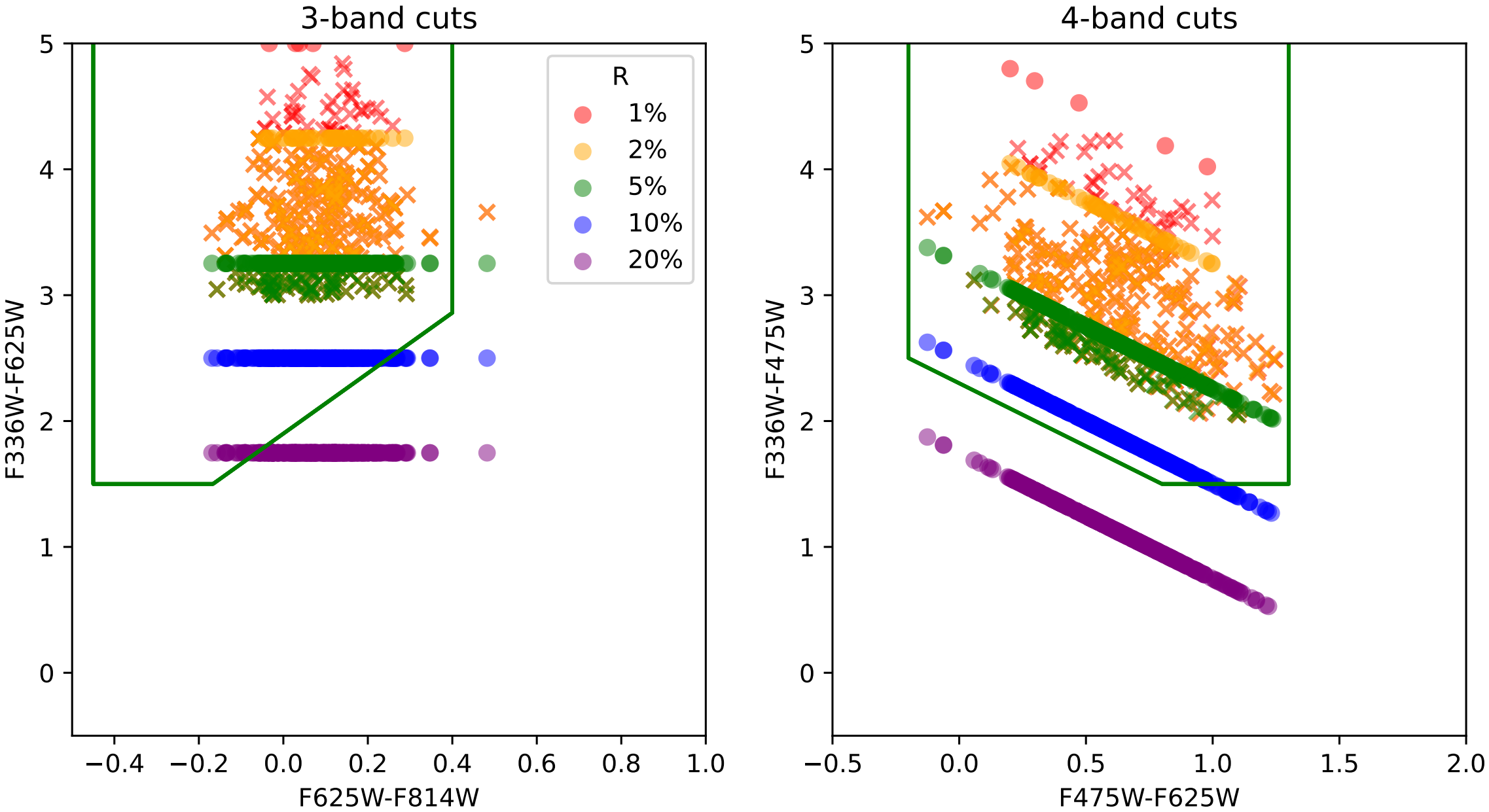

**Figure 10.** The effect of adding LyC emission to our mock galaxies as described in Section 6.2. Left panel: the color selection using the three PIE filters, as in the top row of Figure 8. Right panel: the color selection utilizing the F475W PIE+ data, as shown in the bottom row of Figure 8. We show JAGUAR galaxies that lie at $3.1 < z < 3.5$, with F336W flux added corresponding to different values of $R_{\rm obs}$, denoted by the different colors. Points show galaxies that would be detected in F336W, while crosses show points for which only upper limits would be measured. Points lying outside the region bounded by the green line would not be selected for spectroscopic follow-up.

We show this more directly in Figure 11. We assume a range of possible stellar ages for each of the JAGUAR galaxies at $3.1 < z < 3.5$ (shown by the differently colored lines), and use Equation (7) from J. Chisholm et al. (2019) to convert this to the intrinsic flux ratio between the LyC and UV continuum (older galaxies have a weaker intrinsic LyC). We also assume a range of escape fractions (shown on the $x$-axis), and randomly draw 20 different IGM sight lines for each galaxy from the distributions generated in Section 4.3, allowing us to model the observed $F_{\rm LyC}/F_{\rm UV}$. By adding this LyC flux to the galaxy, we can measure new colors (as in Figure 10) and determine what fraction of galaxies remain inside our selection window for a given age and $f_{\rm esc}$.

These results are shown in Figure 11, illustrating that the completeness of our selection remains high for most values of age and escape fraction. Therefore, the chance of an LyC leaker being rejected from either of our color cuts due to this additional F336W flux is very low, requiring simultaneously a very young stellar age, a very high $f_{\rm esc}$, and an unusually transparent IGM sight line. For example, even a galaxy with a stellar age of 5 Myr and $f_{\rm esc} \approx 100\%$ still requires a sight line among the one-third most transparent to be shifted out of our selection window. With an escape fraction ≈30%, there is <3% chance that such a galaxy would be rejected from our color cuts.

For comparison, the distribution of $R_{\rm obs}$ values measured by C. C. Steidel et al. (2018), at $2.8 \lesssim z \lesssim 3.3$, suggests that only ≈5% of LBGs have $R_{\rm obs} > 0.1$. Drawing from this distribution to obtain LyC flux, and adding this to the JAGUAR mock galaxies, leads to a prediction that <1% of LBGs will have sufficient LyC emission to move out of our selection window. This represents up to 10 galaxies across the entire survey that would not be targeted due to this excess emission. The actual number is likely to be lower, as our galaxies are at slightly higher redshift, so are likely to have a less-transparent IGM sight line.

### *6.3. Selected Targets*

Here we describe the process by which we take our photometric measurements described in Section 5, and produce a galaxy catalog and a list of selected candidate LBGs that we will target with our follow-up spectroscopy.

We apply the above color cuts to the galaxies for which we have measured photometry in all three/four bands. Due to the PIE+ data often having a different orientation, the overlapping area is often substantially smaller than the WFC3 field of view. We exclude sources that are not covered by all three PIE bands. However, we apply the three-band selection to sources in fields that have F475W data and are not covered due to this different orientation.

We first use the "CLASS STAR" parameter measured by SExtractor to remove those sources most likely to be stars. As we are likely to target compact sources, we only remove those for which "CLASS STAR" > 0.8, denoting >80% probability that the source is a star. We then apply the ≈0.2 mag aperture corrections discussed in Section 5.2. As with the mock galaxies, we only consider galaxies with $m$(F625W) < 26.0, in order to ensure S/N > 5 so that we can measure accurate photometry. This also improves the likelihood of successful spectroscopic confirmation.

We then measure the colors of these galaxies, correcting for Milky Way extinction using dust maps from E. F. Schlafly & D. P. Finkbeiner (2011) and the extinction law given by K. D. Gordon et al. (2023); this correction alters our colors by <0.1 mag in most of our fields. The estimated extinction in each band for each field is provided in Appendix B (Table 7). We also identify sources that do not reach a $1\sigma$ level in F336W, and measure the upper limit on their brightness.

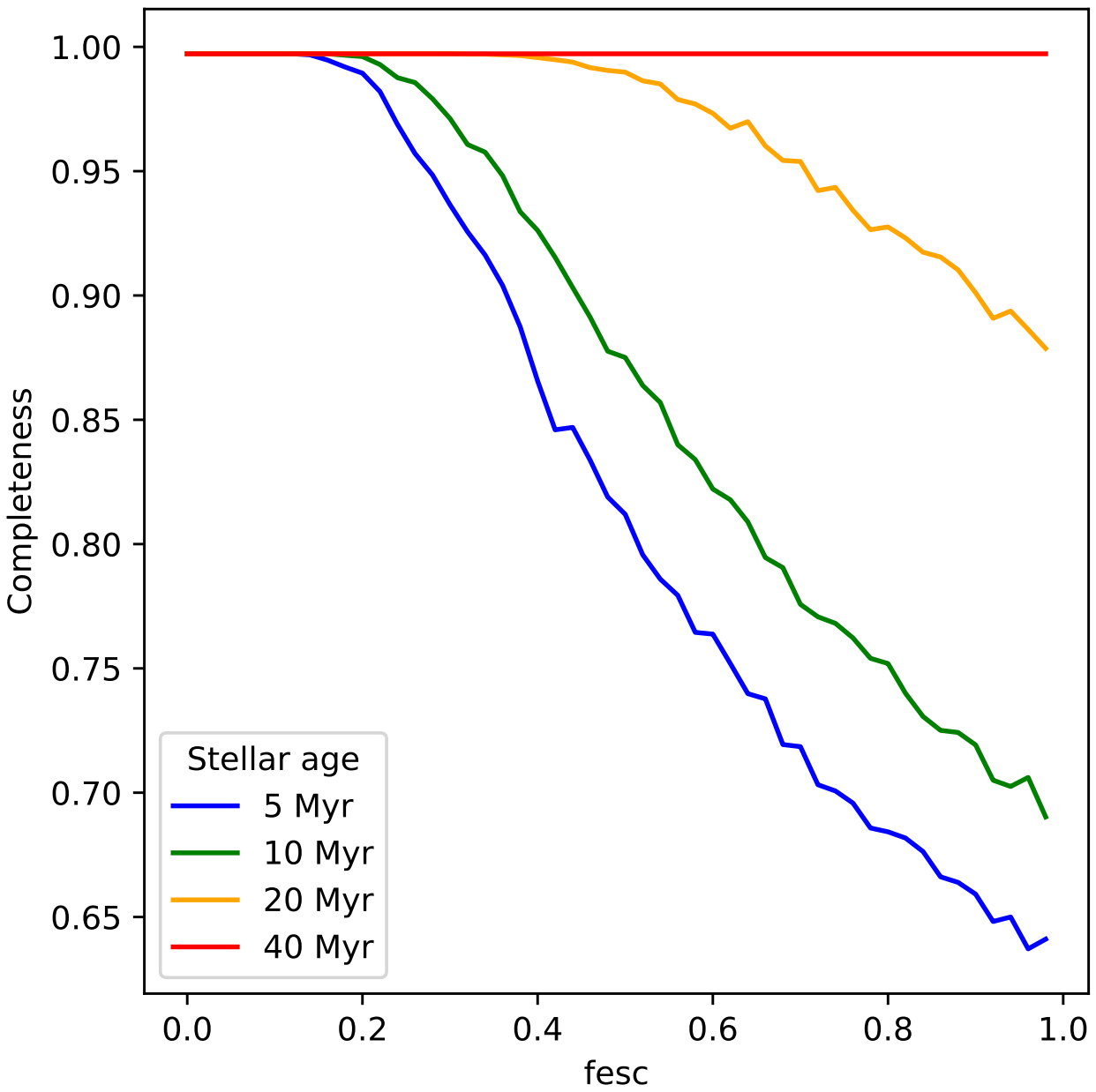

**Figure 11.** The fraction of $3.1 < z < 3.5$ galaxies that will be selected in both of our color cuts as a function of stellar age and $f_{\rm esc}$, assuming the distribution of IGM opacities discussed in Section 4.3. Thus, a completeness of 0.9 indicates that there is a 10% chance that a galaxy with the shown escape fraction and stellar age would be shifted outside of our color selection and missed from the follow-up spectroscopic survey.

The corrected colors for these galaxies are shown in Figure 12. Applying the same color selections optimized using the mocks allows us to select our sample of target galaxies, highlighted in the figure. We also show the magnitude distributions of our sources (aperture-corrected, but not extinction-corrected) in Figure 13. We note that our final target sample removes the likely stars and also excludes sources brighter than $m$(F625W) = 23.5.

In total, we find ≈13,000 sources that are classified as likely galaxies with 23.5 < $m$(F625W) < 26.0, of which ≈1400 fall within our color selection window (or both windows for fields with F475W imaging), and are likely to be LBGs. The number of selected targets in individual fields varies substantially, from ≈10 to ≈60, due to a combination of the varying depths of our detection images and the actual source densities of the different fields. About half of these galaxies are detected at 3$\sigma$ in F336W. Of the ≈1400 LBGs selected, and assuming the 30%–40% purity found in Section 6, we expect ≈500 galaxies that can be stacked in order to measure their LyC emission. Of these, assuming an $R_{\rm obs}$ distribution similar to C. C. Steidel et al. (2018), we predict ≈80 individual LyC emitters to be detected in F336W.

In order to facilitate further work in the community, we will release the data products and catalogs described in this paper on MAST as a High Level Science Product. This will be accessible via the doi:10.17909/emsa-yj13.[16] This data release includes the final images for each field and filter, as well as the full source catalog and a list of galaxies that we aim to target for spectroscopic follow-up (lying in our color selection window and in the 23.5 < $m$(F625W) < 26.0 mag range). We show a sample from the catalog of target galaxies in Table 5.

[16] The HLSP release can also be accessed at https://archive.stsci.edu/hlsp/pie/.

### 6.4. Selection Biases

The color selection in Section 6.1 suggests that ≈30%–40% of the galaxies selected as LBGs will lie in the target redshift range for which F336W probes only LyC emission. However, depending on the telescope/instrument/strategy used for the spectroscopic follow-up, it is not always possible to place slits on all of the candidate LBGs in a field. This will lead to some biases that can affect the resulting sample, which we discuss here. Some of these effects are also described in R. Bassett et al. (2022).

First, we may target selected galaxies with measured flux in F336W, to capture any individual LyC emitters. However, a large fraction of our selected galaxies lie at $z < 3.1$, where F336W flux may lie redward of the Lyman limit. Prioritizing galaxies with detected F336W flux will therefore bias our sample toward these galaxies, reducing the number of galaxies with confirmed redshifts at $3.1 < z < 3.5$. Using the measured $R_{\rm obs}$ values from C. C. Steidel et al. (2018), we predict that only 15%–25% of the selected sources that are also detected in F336W will lie in this target redshift range, significantly lower than the 30%–40% rate for all selected sources. This effect will likely be stronger in the PIE sample, as our galaxies lie at higher redshift than the C. C. Steidel et al. (2018) sample, so fewer LyC leakers will have a sufficiently transparent IGM sight line to show emission. Therefore, in order to ensure we obtain a large sample of galaxies that we can stack in several bins of different galaxy properties, we need to sample F336W nondetections that are more likely to lie at the correct redshifts in addition to the detected galaxies that might be LyC emitters.

Another source of bias may arise from preferentially selecting galaxies that are substantially brighter than the $m = 26$ limit shown in these figures, in order to improve the chances of spectroscopic redshift confirmation. This will again bias our sample toward lower-redshift galaxies, where a larger fraction of galaxies are bright enough to be targeted. We show this bias in Figure 14. Using a brighter magnitude threshold does not noticeably reduce the completeness of our color selection in the target redshift range, although we do become less complete at higher redshifts. This is likely due to probing only the highest-mass galaxies at these redshifts. The fraction of selected sources that lie in the range $3.1 < z < 3.5$ is not substantially affected by limiting our sample to $m$(F625W) < 25.0 or 25.5, with 30%–35% of selected sources lying in this redshift range. This drops to ≈25% when we limit the sample to sources brighter than $m$(F625W) = 24.5. Although our selected sample is weighted toward lower redshifts when using a brighter cut (as expected given the few galaxies brighter than these cuts at high-$z$), the fraction of galaxies in this target range remains the same.

## 7. Spectroscopic Follow-up

Although we now have a large catalog of candidate LBGs, some of which have F336W emission that may indicate LyC escape, confirming that these are indeed LyC leakers requires spectroscopic redshift confirmation. In order to obtain follow-up spectra of our target galaxies, we are utilizing a variety of ground-based instruments including the Low Resolution Imaging Spectrometer (LRIS; J. B. Oke et al. 1995) on the Keck telescope and Binospec on the MMT. These spectra will be used to confirm galaxy redshifts and measure other galaxy properties that can be used to trace LyC emission. Our spectroscopic campaign is ongoing, so these observations will be presented in full in future work. Here we briefly show our first example of spectroscopic confirmation using LRIS.

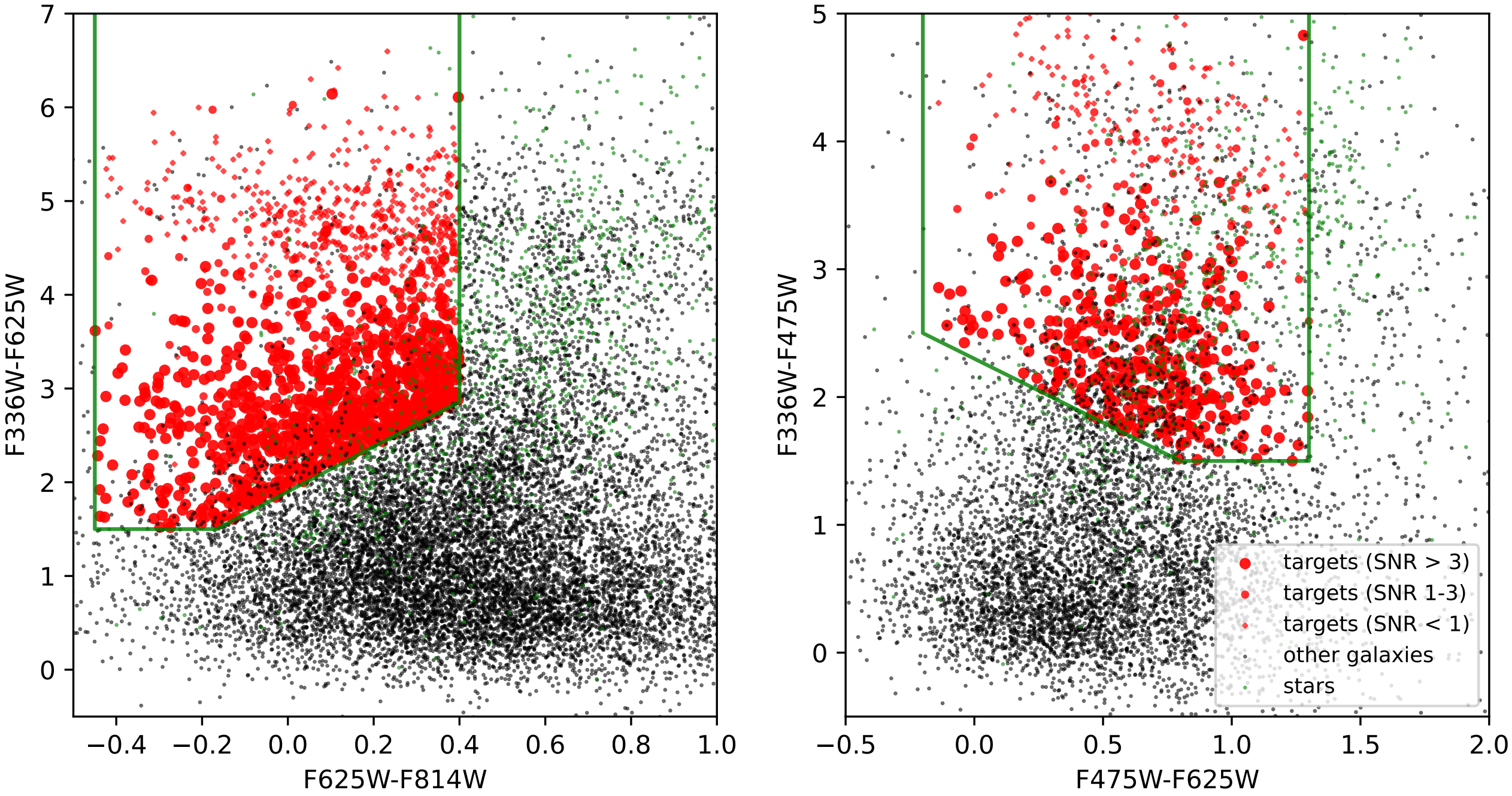

**Figure 12.** Color selection diagrams of sources detected in the PIE fields (left panel: three-band cuts; right panel: additional cuts for fields with F475W). Our selection window is outlined in green in both panels. Target galaxies (based on both color cuts where available) are shown as red points, sized depending on the signal-to-noise in the F336W image. Other galaxies are black points, and stars are shown in green.

### 7.1. LRIS Observations

We have observed seven masks across four fields (fields 9, 22, 34, and 35) using LRIS between 2023 October and 2024 January, under program U127 (PI: Malkan). We use the grating and *grism* with $R \sim 600$, with the *grism* centered on 4000 Å in the blue channel and the grating centered on 7500 Å in the red channel, split using the dichroic at 5600 Å. The blue channel therefore covers 3400–5600 Å, ensuring that Ly$\alpha$ is covered across our target redshift range. The red channel extends up to $\approx$9000 Å, potentially allowing us to detect a variety of UV emission lines from species such as Si IV, C IV, He II, and C III]. Low-redshift interlopers may be identified with emission lines such as [O II] (3727 Å), H$\beta$, [O III] (4959, 5007 Å), and H$\alpha$.

For most masks, we used 1.″2 slits, but switched to 1.″5 slits when the seeing was worse than 1.″2 in order to minimize flux losses. Due to observational constraints, exposure times vary between the different masks, resulting in total exposure times between 2400 and 18,000 s on each source. Arc, dome flat, and standard star exposures were taken on each night with the same configuration.

The LRIS data are reduced using PypeIt (J. Prochaska et al. 2020). The locations of the slit edges are traced using the dome flats. The arc frames are combined into a "master" arc file, which is then used for wavelength calibration; a master flat is similarly used to correct the science exposures. The 2D spectra are coadded and then extracted using a boxcar to produce 1D spectra for each source. These are then flux-calibrated using the spectra of the standard star.

However, we note that the reduction efforts for this data are still ongoing. In particular, the increased number of sky lines in the wavelength range covered by the red arm makes the processing more complex, so this is not yet complete, and the data from the red arm cannot currently be used to help constrain the redshifts of our sources.

We show an example spectrum from the blue arm of LRIS in Figure 15. The photometry of this object is given in Table 5. It lies in field 9, and was observed for a total of 10,200 s with a 1.″2 slit. The strong emission line makes the redshift measurement more straightforward than for other sources. We identify this line as Ly$\alpha$ at $z \approx 3.0$. Noting that inflows and outflows can shift the apparent Ly$\alpha$ peak away from the systemic redshift (e.g., A. Verhamme et al. 2018), we use the absorption features identified as Ly$\beta$ and Ly$\delta$ in the lower panels of Figure 15 to measure a systemic redshift of $z = 2.987$. As discussed above, we expect a large fraction of our targeted galaxies to lie in the $2.7 < z < 3.1$ range.

Although we cannot rule out that this might be a low-redshift [O II] emitter, we do not see any absorption features redward of the emission line (such as the Ca *K* and *H* bands) at the wavelengths where they might be expected if this were a low-redshift galaxy.

A small amount of flux redward of the Lyman limit still falls in the F336W filter, so we cannot directly measure the LyC emission from this galaxy using the HST imaging. However, our measured magnitude of 28.15 (stated in Table 5) does represent an upper limit on the LyC flux. This is only a $\approx 2\sigma$ detection, so the source is not clear in the F336W postage-stamp image.

The Keck spectrum does appear to show some flux blueward of the Lyman limit, so we can integrate the Keck spectrum to provide a lower limit on the LyC flux. As shown in Figure 15, the region between 890 and 910 Å (rest-frame) is fairly clean, lying between a strong sky line and the Lyman limit. Integrating this region yields a total flux of $3.5 \times 10^{-17}$ ergs s$^{-1}$ cm$^{-2}$, which would be measured as magnitude $\approx$28.2 in the F336W filter.

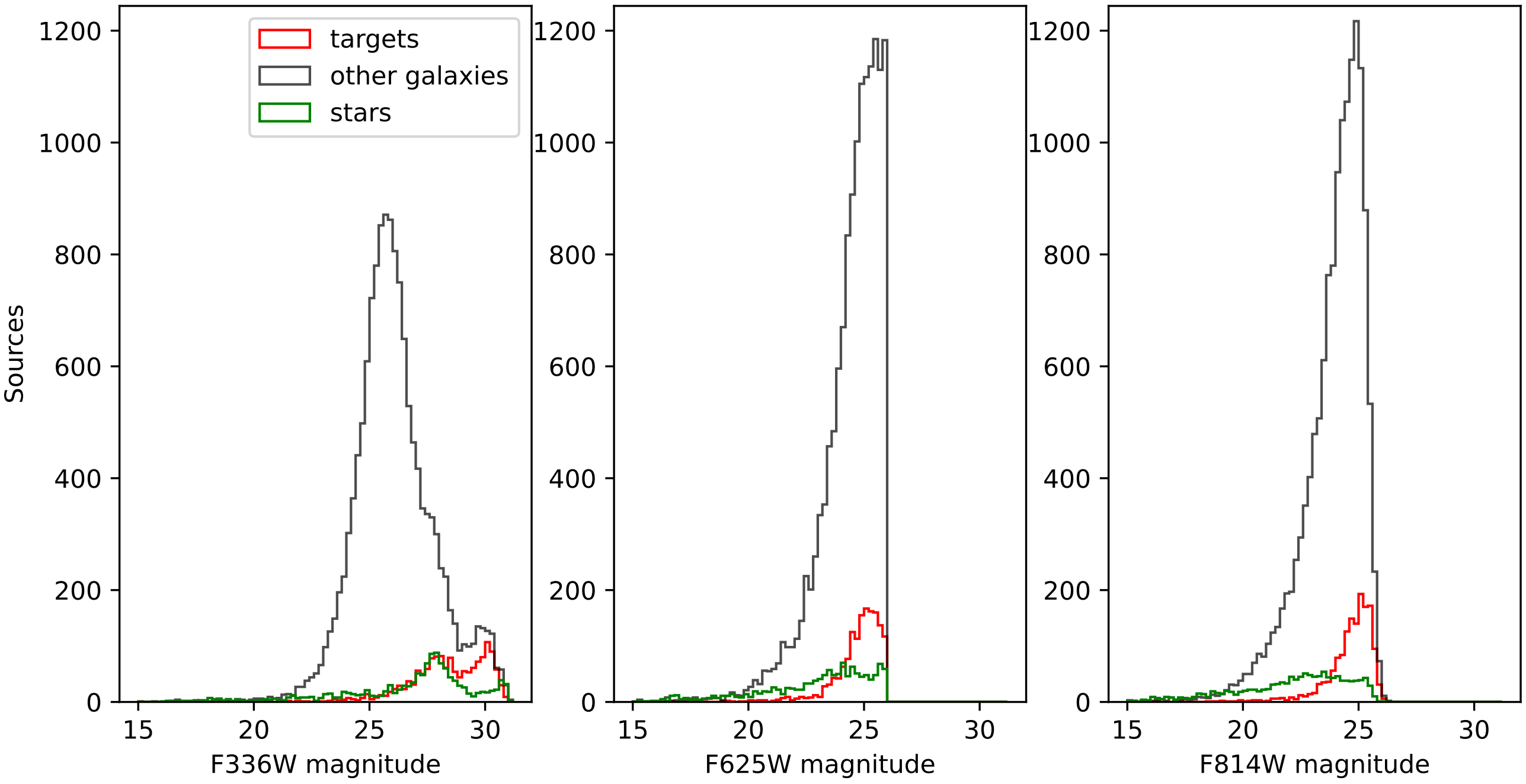

**Figure 13.** Observed magnitude distribution of detected sources in each PIE filter with $m$(F625W) $< 26$. We show sources classified as stars, and those selected by our color cuts, showing that stars are more likely to be brighter than galaxies, and most of our color-selected galaxies are fainter than $m$(F625W) $= 23$, consistent with our mock predictions.

**Table 5**
First 10 Rows of the Target Catalog Described in Section 6.3, Providing Target Locations and Magnitudes for All Observed Bands

| ID | R.A. (deg) | Decl. (deg) | Field | $m$(F336W) (mag) | $m$(F475W) (mag) | $m$(F625W) (mag) | $m$(F814W) (mag) | CLASS STAR |
|---|---|---|---|---|---|---|---|---|
| PIE-J103700.95+371028.96 | 159.2539739 | 37.174713 | 1 | 26.69 ± 0.11 | ⋯ | 25.09 ± 0.09 | 25.46 ± 1.13 | 0.06 |
| PIE-J103710.43+371118.05 | 159.2934634 | 37.188348 | 1 | >30.33 | ⋯ | 25.70 ± 0.06 | 25.41 ± 0.22 | 0.00 |
| PIE-J103700.61+371005.37 | 159.2525772 | 37.168159 | 1 | 27.97 ± 0.13 | ⋯ | 25.21 ± 0.04 | 25.27 ± 0.18 | 0.00 |
| PIE-J103708.12+371052.72 | 159.2838654 | 37.181312 | 1 | >30.47 | 26.33 ± 0.09 | 25.26 ± 0.04 | 25.16 ± 0.18 | 0.00 |
| PIE-J103708.35+371038.88 | 159.2848327 | 37.177468 | 1 | 29.26 ± 0.34 | 26.57 ± 0.07 | 25.50 ± 0.05 | 25.54 ± 0.21 | 0.01 |
| PIE-J103712.89+371108.25 | 159.3037333 | 37.185626 | 1 | >30.31 | ⋯ | 25.56 ± 0.07 | 25.36 ± 0.33 | 0.00 |
| PIE-J103713.12+371102.95 | 159.3046982 | 37.184155 | 1 | 27.19 ± 0.07 | ⋯ | 25.30 ± 0.04 | 25.55 ± 0.26 | 0.01 |
| PIE-J103713.22+371101.52 | 159.3051247 | 37.183758 | 1 | 28.56 ± 0.24 | ⋯ | 24.93 ± 0.03 | 24.97 ± 0.16 | 0.00 |
| PIE-J103711.01+371045.79 | 159.295889 | 37.179386 | 1 | >30.06 | 25.01 ± 0.03 | 24.64 ± 0.03 | 24.35 ± 0.13 | 0.00 |
| PIE-J103707.60+371022.99 | 159.281669 | 37.173054 | 1 | 28.49 ± 0.14 | 26.48 ± 0.06 | 25.89 ± 0.10 | 26.03 ± 0.26 | 0.03 |
| PIE-J083655.92+191813.59 | 129.233006 | 19.303781 | 9 | 28.15 ± 0.24 | 25.51 ± 0.04 | 24.82 ± 0.04 | 24.92 ± 0.17 | 0.00 |

**Note.** The full catalog will be released on MAST. We also show the galaxy discussed in Section 7. We include the first 10 objects from field 1, as well as the two objects from field 22 for which example spectra are shown in Section 7.

We can also use the F475W observations to verify the flux calibration. The blue-arm LRIS spectrum integrates to provide a predicted F475W magnitude of ≈25.3, compared to the $m$(F475W) ≈ 25.5 observed in the HST image of this source, a difference of ≈20% in flux. We note that we have not yet compared the measured fluxes with existing source catalogs to better verify this uncertainty on our flux measurements. Alongside the low signal-to-noise from the F336W image, there is still a large uncertainty (>30%) on these LyC constraints.

An LyC measurement of $m$(F336W) ≈ 28.2 corresponds to $R_{\mathrm{obs}} \approx 0.04$, given the measured F625W flux from this galaxy in the HST image. Estimating an LyC escape fraction from this requires measuring or assuming an intrinsic LyC/UV flux ratio and the IGM transmission $T_{\mathrm{IGM}}$. If we assume the median $T_{\mathrm{IGM}}$ at a similar redshift ($T_{\mathrm{IGM}} = 0.07$ at $z = 3.1$, as found in Section 4.3), and a stellar population that is 10 Myr old (using the relation provided by J. Chisholm et al. 2019, and discussed in Section 6.2 to convert this to an intrinsic flux ratio), we can estimate a relative escape fraction $f_{\mathrm{esc}} \approx 60\%$. Both of these assumptions are highly uncertain. Future PIE papers will utilize the fully reduced spectra to better estimate the intrinsic flux ratio, and use our large sample of galaxies to average over the $T_{\mathrm{IGM}}$ distribution to obtain better estimates of $f_{\mathrm{esc}}$ as a function of galaxy properties.

We also note that we currently cannot rule out the possibility that this galaxy is a low-redshift [O II] emitter. Reducing the red arm of LRIS will allow us to discriminate between Ly$\alpha$ and [O II] emission for this and other similar emission-line galaxies. Future work will present the full spectra obtained from these Keck observations, as well as spectroscopy from the MMT for a small number of fields. We will also obtain spectra from JWST (program 9365, PI: Scarlata) covering optical emission lines such as H$\beta$ and [O

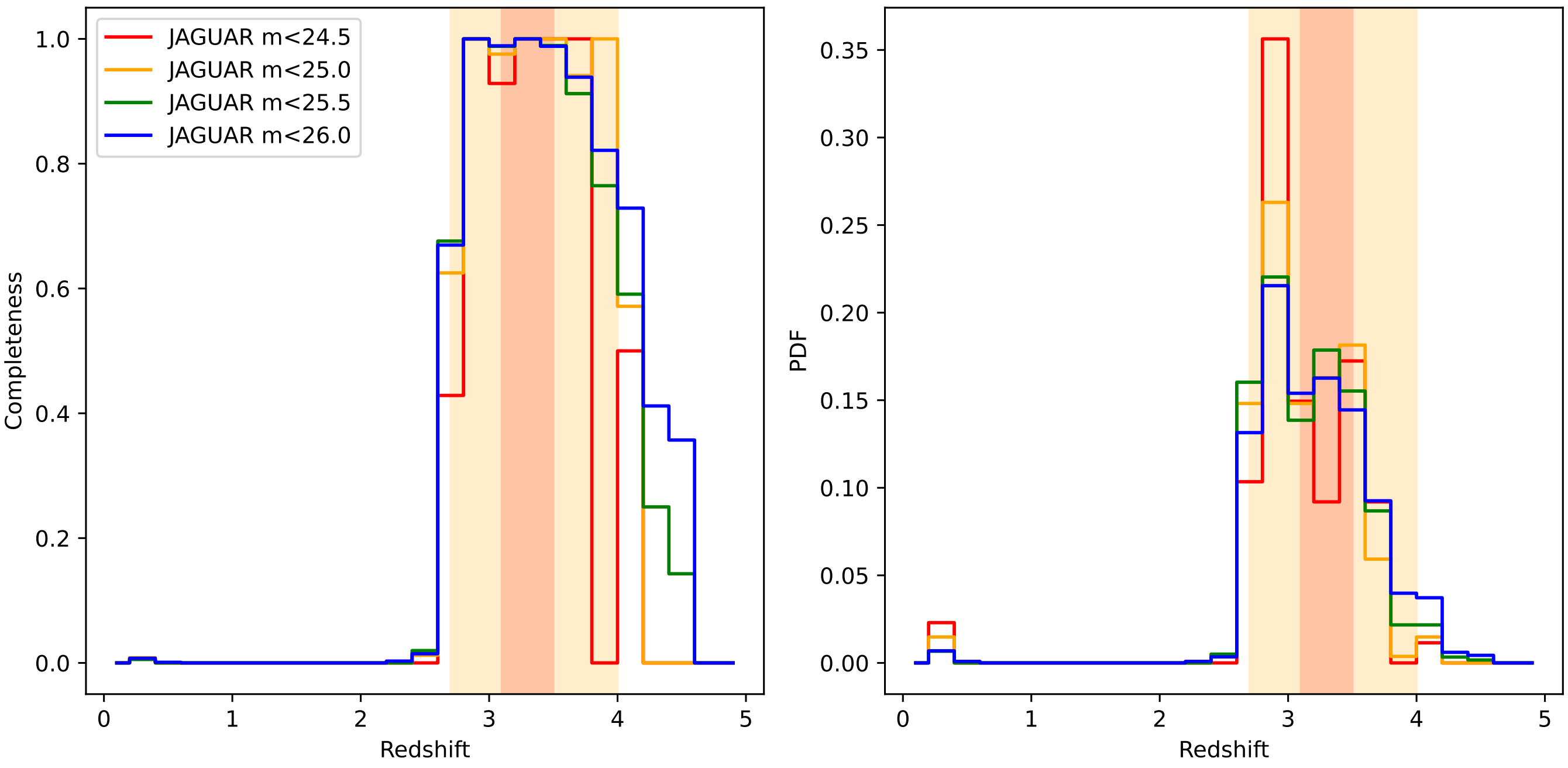

**Figure 14.** Redshift dependence of galaxies selected by the color cuts shown in Figure 8, where we vary the optical (F625W) magnitude required for a galaxy to be included. These are shown in redshift bins of $\Delta z = 0.2$. Left panel: fraction of galaxies in each redshift bin that are selected by the color cuts. Right panel: fraction of galaxies selected by the color cuts that lie in each redshift bin. Shaded regions in both panels represent the $2.7 < z < 4.0$ range for which we optimize the three-band cuts (yellow), and the $3.1 < z < 3.5$ redshift range that we target with the four-band cuts (red).

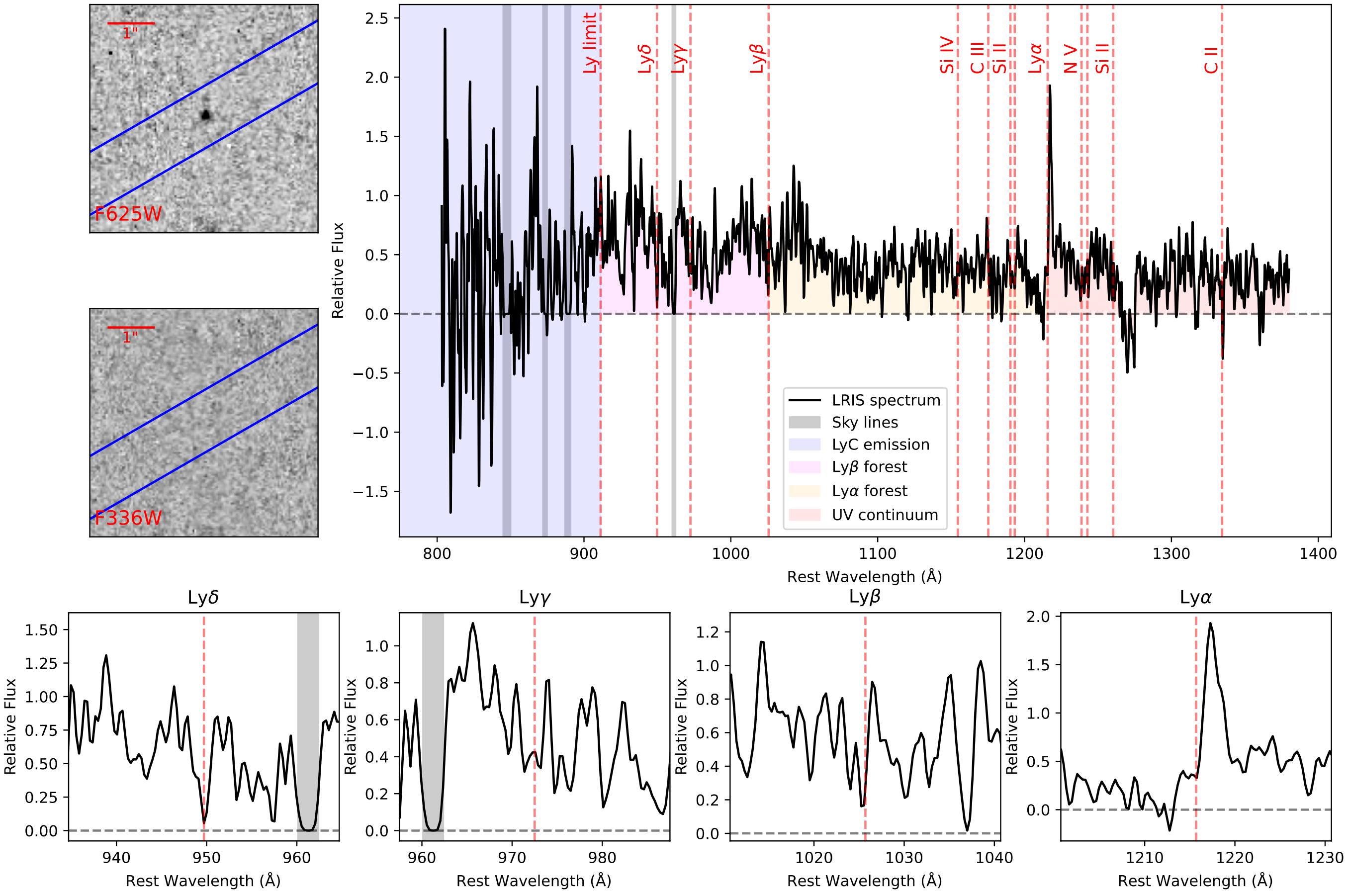

**Figure 15.** Example spectrum from our Keck/LRIS observations, of a galaxy at $z = 2.987$. The spectrum is smoothed using a Gaussian with 1 pixel standard deviation. The spectral ranges covered by the Ly$\beta$ and Ly$\alpha$ forests, and the UV continuum are colored. The left panels show the HST F625W and F336W imaging of this source, with the slit location marked in blue. The lower panels show zoomed-in regions of the spectra around some spectral features. Regions that are affected by strong sky residuals are zeroed-out and shaded in gray.

III] (5007 Å) at $z \sim 3$ for 12 of the PIE fields. This early example spectrum from Keck is included here to help demonstrate that our color selection does allow us to detect $z \approx 3$ galaxies and that their redshifts can be confirmed with ground-based spectrographs.

## 8. Summary and Conclusions

In this paper, we present the design of the PIE survey, an HST program aimed at selecting a large sample of galaxies at $3.1 < z < 3.5$ in order to measure their LyC emission, and characterize this emission as a function of other galaxy properties. We describe and validate our photometric measurements from the HST data and our criteria for selecting targets for follow-up spectroscopy. To summarize, we find the following:

1. Simulated images show that we can accurately recover galaxy colors from the HST data, with systematic uncertainties lower than the image noise. Recovering accurate magnitudes requires an aperture correction of $\approx$0.2 mag, as well as extinction corrections that are usually $\lesssim$0.1 mag.
2. Color selection using the three filters that make up the core PIE survey (F336W, F625W, and F814W) allows us to reliably select $2.7 < z < 4.0$ galaxies. According to our simulations, $\gtrsim$90% of our selected galaxies will be LBGs at this redshift. However, only $\approx$30% of these lie in the $3.1 < z < 3.5$ range for which we can measure LyC emission using the F336W data.
3. The use of F475W allows us to more accurately select for galaxy redshifts. With this filter, $\approx$40% of our sample lies in our target redshift range.
4. We select $\approx$1400 galaxies from the PIE fields that we can target for spectroscopic follow-up, of which we expect $\approx$500 to lie at $3.1 < z < 3.5$ (suitable for stacking) and $\approx$80 detections of individual LyC emitters.
5. We show spectroscopic redshift confirmation using Ly$\alpha$ emission for our first targeted galaxy using Keck/LRIS (mag 25 galaxy observed for $\approx$10,000 s), demonstrating this will be practical for a large fraction of our selected galaxies. Fluxes of emission lines such as Ly$\alpha$ are candidates for indirect LyC tracers that can be used in the epoch of reionization (although recent works suggest Ly$\alpha$ may be a poor proxy for LyC escape).
6. The transmission of the IGM varies substantially across the sky, and can be correlated on large scales. The use of independent fields will allow us to use the average IGM transmission when stacking LBGs, and reduce the uncertainty on our measurements of LyC escape.

Future work will present spectroscopic data of the PIE fields, including Keck, MMT, and JWST spectroscopy, as well as morphologies of individual LyC emitters, and LyC escape fractions as a function of galaxy properties and emission-line diagnostics. These diagnostics, if found to trace LyC emission in the local Universe and at $z > 3$, will allow us to determine to what extent galaxies contributed to the reionization of the Universe, and which galaxies dominated this contribution.

## Acknowledgments

We thank the anonymous reviewer for providing thoughtful comments that have improved this manuscript.

This work is based on observations with the NASA/ESA Hubble Space Telescope obtained from the MAST Data Archive at the Space Telescope Science Institute, which is operated by the Association of Universities for Research in Astronomy, Incorporated, under NASA contract NAS5-26555. In particular, this work is based on data from programs 17147 and 17518. All of the data from these programs can be accessed via doi:10.17909/dww5-0e41, and processed images and catalogs will be made available at doi:10.17909/emsa-yj13. This research has also made use of NASA's Astrophysics Data System.

M.J.H. is fellow of the Knut & Alice Wallenberg Foundation. A.J.B. acknowledges funding from the "FirstGalaxies" Advanced Grant from the European Research Council (ERC) under the European Union's Horizon 2020 research and innovation program (grant agreement No. 789056).

Some of the data presented herein were obtained at Keck Observatory, which is a private 501(c)3 nonprofit organization operated as a scientific partnership among the California Institute of Technology, the University of California, and the National Aeronautics and Space Administration. The Observatory was made possible by the generous financial support of the W. M. Keck Foundation. The authors wish to recognize and acknowledge the very significant cultural role and reverence that the summit of Maunakea has always had within the Native Hawaiian community. We are most fortunate to have the opportunity to conduct observations from this mountain.

*Facilities:* HST (WFC3), Keck:I (LRIS).

*Software:* Astropy (Astropy Collaboration et al. 2013, 2018, 2022), DustExtinction (K. Gordon 2024), IPython (F. Perez & B. E. Granger 2007), Jupyter (T. Kluyver et al. 2016), Matplotlib (J. D. Hunter 2007; T. A. Caswell et al. 2023) Numpy (C. R. Harris et al. 2020), Photutils (L. Bradley et al. 2024) Scipy (P. Virtanen et al. 2020), SExtractor (E. Bertin & S. Arnouts 1996).

## Appendix A
## Unusable Data

For completeness, and to facilitate future use of the PIE data, here we describe the data produced by the PIE and PIE+ programs that could not be used to contribute to the survey. As mentioned in the text, several fields were affected by poor-quality data, most often due to the HST gyro issues affecting many Cycle 30 observations. In other cases, the observations did not execute at all. The data for fields that do not have at least three filters of usable data are not included in the MAST release of science products, but can be accessed through the archive portal. In Table 6, we show the usable and total exposure times for each field and filter. Some images were not used in the PIE survey due to other missing data; these are marked as usable in this table.

**Table 6**
Description of the Usable and Unusable Observations for Each PIE Field

| Field | R.A. | Decl. | PIE F336W | | PIE F625W | | PIE F814W | | PIE+ Filter | PIE+ | |
|---|---|---|---|---|---|---|---|---|---|---|---|
| | hms | dms | Usable (s) | Total (s) | Usable (s) | Total (s) | Usable (s) | Total (s) | | Usable (s) | Total (s) |
| 01 | 10:37:09.642 | +37:09:30.66 | 8209 | 8209 | 1150 | 1150 | 1150 | 1150 | F475W | 1500 | 1500 |
| 02 | 12:56:31.700 | –05:45:23.00 | 4142 | 4142 | 1150 | 1150 | 1120 | 1120 | … | … | … |
| 03 | 10:17:30.526 | +03:46:23.82 | 1560 | 1560 | 2020 | 2020 | 0 | 0 | F475W | 1500 | 1500 |
| 04 | 13:43:11.505 | –00:51:27.41 | 16837 | 16837 | 2043 | 2043 | 2046 | 2046 | … | … | … |
| 05 | 09:56:21.151 | +28:48:06.93 | 10477[a] | 14866 | 2276 | 2276 | 2264 | 2264 | F475W | 1500 | 1500 |
| 06 | 10:42:05.548 | +18:20:52.16 | 6259 | 8149 | 1150 | 1150 | 1150 | 1150 | F475W | 1500 | 1500 |
| 07 | 14:34:55.651 | +20:10:06.92 | 8156 | 8156 | 1150 | 1150 | 1130 | 1130 | … | … | … |
| 08 | 10:21:22.942 | +18:03:46.79 | 15611 | 15611 | 2070 | 2070 | 2370 | 2370 | F475W | 3000 | 3000 |
| 09 | 08:37:01.250 | +19:18:39.05 | 5906 | 8430 | 1150 | 1150 | 1150 | 1150 | F475W | 1500 | 1500 |
| 10 | 08:15:31.114 | +29:40:03.50 | 6960 | 8852 | 1170 | 1170 | 0 | 1170 | F814W | 0 | 1500 |
| 11 | 10:03:44.065 | +32:56:59.70 | 0 | 8549 | 0 | 1150 | 0 | 1150 | … | … | … |
| 12 | 09:49:27.882 | +48:30:50.28 | 7442 | 7442 | 1158 | 2259 | 0 | 0 | F814W | 1500 | 1500 |
| 13 | 12:42:51.863 | +36:31:14.73 | 2070[b] | 3170 | 1182 | 1182 | 0 | 0 | F814W | 1500 | 1500 |
| 14 | 09:33:37.868 | +55:10:12.63 | 15560 | 15560 | 1114 | 2228 | 2280 | 2280 | F475W | 1500 | 3000 |
| 15 | 12:47:37.246 | +58:19:56.88 | 3490 | 3490 | 1101 | 1101 | 0 | 0 | F814W | 1500 | 1500 |
| 16 | 15:27:01.362 | –23:31:45.17 | 27779 | 34911 | 4788 | 4788 | 2400 | 2400 | … | … | … |
| 17 | 11:45:12.397 | +62:03:40.81 | 7880 | 7880 | 1047 | 1047 | 1152 | 1152 | F475W | 1500 | 1500 |
| 18 | 13:05:21.508 | +52:57:20.30 | 2700[b] | 7648 | 1047 | 1047 | 1152 | 1152 | F475W | 1500 | 1500 |
| 19 | 12:12:35.331 | +61:48:24.45 | 5082 | 7832 | 0 | 1047 | 0 | 1152 | … | … | … |
| 20 | 14:18:51.037 | +42:03:36.75 | 6225 | 7528 | 2259 | 2259 | 0 | 0 | F814W | 0 | 1500 |
| 21 | 15:04:50.249 | +37:17:34.56 | 5148 | 8881 | 0 | 1110 | 780 | 1170 | … | … | … |
| 22 | 22:29:31.974 | +27:30:54.07 | 8590 | 8590 | 1150 | 1150 | 1130 | 1130 | F475W | 1500 | 1500 |
| 23 | 14:27:19.886 | +26:28:02.71 | 6763 | 9293 | 1170 | 1170 | 1170 | 1170 | F475W | 0 | 1500 |
| 24 | 16:07:27.158 | +32:14:01.81 | 6600 | 6600 | 2262 | 2262 | 0 | 0 | … | … | … |
| 25 | 13:40:12.756 | +54:43:38.30 | 9470 | 9470 | 1200 | 1200 | 1200 | 1200 | F475W | 1500 | 1500 |
| 26 | 15:44:17.383 | +27:38:22.94 | 9273 | 9273 | 1170 | 1170 | 1170 | 1170 | … | … | … |
| 27 | 16:37:57.174 | +33:44:46.39 | 8801 | 8801 | 1150 | 1150 | 1120 | 1120 | … | … | … |
| 28 | 09:09:41.478 | +27:39:43.39 | 5995 | 5995 | 2020 | 2020 | 0 | 0 | F475W | 1500 | 1500 |
| 29 | 10:05:38.081 | +60:27:36.66 | 5845 | 8165 | 1150 | 1150 | 1130 | 1130 | F475W | 0 | 1500 |
| 30 | 07:45:47.488 | +19:22:54.40 | 6854 | 6854 | 2208 | 2208 | 0 | 0 | F475W | 1500 | 1500 |
| 31 | 09:27:42.579 | +30:47:27.31 | 8140 | 8140 | 1120 | 1120 | 1130 | 1130 | F475W | 1500 | 1500 |
| 32 | 11:02:01.208 | +51:32:19.40 | 16370 | 16370 | 2004 | 2004 | 2571 | 2571 | F475W | 1500 | 3000 |
| 33 | 12:02:35.358 | +55:08:17.80 | 7675 | 7675 | 1145 | 1145 | 1090 | 1090 | F475W | 1500 | 1500 |
| 34 | 10:16:54.722 | +47:06 17.32 | 21256 | 21256 | 1977 | 1977 | 1977 | 1977 | F475W | 1500 | 1500 |
| 35 | 09:57:10.764 | +28:51:35.55 | 21178 | 21178 | 2300 | 2300 | 2265 | 2265 | F475W | 3000 | 3000 |
| 36 | 20:43:53.591 | –10:39:52.54 | 13974 | 13974 | 2394 | 2394 | 2388 | 2388 | F475W | 1500 | 1500 |
| 37 | 10:11:22.926 | –04:39:54.77 | 17634 | 17634 | 2082 | 2082 | 2382 | 2382 | F475W | 1500 | 1500 |
| 38 | 02:12:23.782 | +00:55:28.22 | 15546 | 15546 | 0 | 2064 | 2382 | 2382 | F625W | 1500 | 1500 |
| 39 | 06:29:01.868 | –50:55:13.50 | 0 | 10290 | 1150 | 1150 | 0 | 1130 | F814W | 1500 | 1500 |
| 40 | 02:25:12.584 | +00:10:31.22 | 2640[b] | 14520 | 2388 | 2388 | 0 | 2379 | F814W | 1500 | 1500 |
| 41 | 12:25:23.770 | +16:26:30.24 | 8231 | 8231 | 1150 | 1150 | 1150 | 1150 | … | … | … |
| 42 | 09:31:52.832 | +32:00:54.65 | 8155 | 8155 | 1145 | 1145 | 1130 | 1130 | … | … | … |
| 43 | 12:20:56.068 | +17:22:05.82 | 8231 | 8231 | 1150 | 1150 | 1150 | 1150 | … | … | … |
| 44 | 09:56:24.891 | +28:53:01.56 | 9080 | 11580 | 2270 | 2270 | 2260 | 2260 | … | … | … |
| 45 | 09:33:48.771 | +51:14:37.96 | 9760 | 9760 | 980 | 980 | 980 | 980 | … | … | … |
| 46 | 09:56:40.324 | +17:35:30.18 | 4000 | 6500 | 1158 | 1158 | 0 | 0 | F814W | 1500 | 1500 |
| 47 | 09:09:09.284 | +39:34:25.88 | 8180 | 8180 | 1020 | 1020 | 1300 | 1300 | F475W | 1500 | 1500 |
| 48 | 12:10:22.266 | +16:47:12.45 | 6259 | 8184 | 1150 | 1150 | 1150 | 1150 | … | … | … |
| 49 | 10:21:20.944 | +04:34:18.84 | 2706[b] | 8804 | 0 | 1120 | 730 | 1120 | F475W | 1500 | 1500 |
| 50 | 09:55:01.504 | +39:34:05.51 | 7896 | 8972 | 1200 | 1200 | 0 | 1200 | F475W | 1500 | 1500 |
| 51 | 10:00:14.228 | +05:23:03.45 | 15601 | 15601 | 2097 | 2097 | 2385 | 2385 | F475W | 1500 | 1500 |
| 52 | 06:27:55.994 | –50:57:10.94 | 8950 | 12900 | 1150 | 1150 | 0 | 1130 | F475W | 1500 | 1500 |
| 53 | 10:03:46.445 | +32:53:48.35 | 8340 | 8340 | 1200 | 1200 | 1300 | 1300 | … | … | … |
| 54 | 00:41:16.052 | –51:26:50.10 | 18298 | 18298 | 0 | 0 | 2334 | 2334 | … | … | … |

**Notes.**
[a] Field 5—One visit consists of only one exposure, which could not be aligned with the other visit due to cosmic rays and, hence, was not included.
[b] Fields 13, 18, 40, and 49—usable F336W too short for image alignment to be confirmed or <3 exposures mean that cosmic rays could not be removed.

## Appendix B
## Extinction

Although we avoid targeting fields near to the galactic plane, extinction caused by dust in the Milky Way is not negligible, and must be accounted for in order to determine accurate colors. As discussed in Section 6.3, we use dust maps from E. F. Schlafly & D. P. Finkbeiner (2011) to determine the extinction $E(B-V)$, which is then translated to an attenuation in each band using the extinction law given by K. D. Gordon et al. (2023),[17] assuming the commonly used value of $R(V) = 3.1$. The resulting corrections are listed in Table 7.

**Table 7**
Extinction Values Adopted for Each of the Usable PIE Fields

| Field | R.A. | Decl. | $E(B-V)$ | $A(x)$ | | | |
|---|---|---|---|---|---|---|---|
| | | | | F336W | F475W | F625W | F814W |
| | hms | dms | (mag) | (mag) | (mag) | (mag) | (mag) |
| 01 | 10:37:09.642 | +37:09:30.66 | 0.012 | 0.059 | 0.042 | 0.030 | 0.021 |
| 02 | 12:56:31.700 | −05:45:23.00 | 0.024 | 0.122 | 0.087 | 0.062 | 0.043 |
| 04 | 13:43:11.505 | −00:51:27.41 | 0.025 | 0.128 | 0.092 | 0.066 | 0.045 |
| 05 | 09:56:21.151 | +28:48:06.93 | 0.016 | 0.079 | 0.056 | 0.040 | 0.028 |
| 06 | 10:42:05.548 | +18:20:52.16 | 0.027 | 0.139 | 0.099 | 0.071 | 0.049 |
| 07 | 14:34:55.651 | +20:10:06.92 | 0.027 | 0.140 | 0.100 | 0.071 | 0.049 |
| 08 | 10:21:22.942 | +18:03:46.79 | 0.022 | 0.113 | 0.081 | 0.058 | 0.040 |
| 09 | 08:37:01.250 | +19:18:39.05 | 0.021 | 0.105 | 0.075 | 0.054 | 0.037 |
| 12 | 09:49:27.882 | +48:30:50.28 | 0.011 | 0.057 | 0.041 | 0.029 | 0.020 |
| 14 | 09:33:37.868 | +55:10:12.63 | 0.030 | 0.153 | 0.110 | 0.078 | 0.054 |
| 15 | 12:47:37.246 | +58:19:56.88 | 0.010 | 0.051 | 0.036 | 0.026 | 0.018 |
| 16 | 15:27:01.362 | −23:31:45.17 | 0.094 | 0.482 | 0.345 | 0.246 | 0.170 |
| 17 | 11:45:12.397 | +62:03:40.81 | 0.019 | 0.095 | 0.068 | 0.049 | 0.033 |
| 22 | 22:29:31.974 | +27:30:54.07 | 0.044 | 0.222 | 0.159 | 0.114 | 0.078 |
| 23 | 14:27:19.886 | +26:28:02.71 | 0.015 | 0.075 | 0.053 | 0.038 | 0.026 |
| 25 | 13:40:12.756 | +54:43:38.30 | 0.010 | 0.047 | 0.034 | 0.024 | 0.017 |
| 26 | 15:44:17.383 | +27:38:22.94 | 0.029 | 0.150 | 0.107 | 0.076 | 0.053 |
| 27 | 16:37:57.174 | +33:44:46.39 | 0.021 | 0.108 | 0.077 | 0.055 | 0.038 |
| 29 | 10:05:38.081 | +60:27:36.66 | 0.011 | 0.056 | 0.040 | 0.028 | 0.020 |
| 31 | 09:27:42.579 | +30:47:27.31 | 0.017 | 0.086 | 0.062 | 0.044 | 0.030 |
| 32 | 11:02:01.208 | +51:32:19.40 | 0.009 | 0.043 | 0.030 | 0.022 | 0.015 |
| 33 | 12:02:35.358 | +55:08:17.80 | 0.011 | 0.055 | 0.040 | 0.028 | 0.019 |
| 34 | 10:16:54.722 | +47:06:17.32 | 0.007 | 0.038 | 0.027 | 0.019 | 0.013 |
| 35 | 09:57:10.764 | +28:51:35.55 | 0.016 | 0.079 | 0.056 | 0.040 | 0.028 |
| 36 | 20:43:53.591 | −10:39:52.54 | 0.048 | 0.247 | 0.177 | 0.126 | 0.087 |
| 37 | 10:11:22.926 | −04:39:54.77 | 0.038 | 0.200 | 0.143 | 0.103 | 0.071 |
| 38 | 02:12:23.782 | +00:55:28.21 | 0.027 | 0.142 | 0.102 | 0.073 | 0.050 |
| 41 | 12:25:23.770 | +16:26:30.24 | 0.021 | 0.104 | 0.075 | 0.053 | 0.037 |
| 42 | 09:31:52.832 | +32:00:54.65 | 0.017 | 0.089 | 0.064 | 0.046 | 0.031 |
| 43 | 12:20:56.068 | +17:22:05.82 | 0.023 | 0.120 | 0.086 | 0.062 | 0.042 |
| 44 | 09:56:24.891 | +28:53:01.56 | 0.017 | 0.101 | 0.072 | 0.052 | 0.036 |
| 45 | 09:33:48.771 | +51:14:37.96 | 0.012 | 0.064 | 0.046 | 0.033 | 0.023 |
| 46 | 09:56:40.324 | +17:35:30.18 | 0.024 | 0.119 | 0.085 | 0.061 | 0.042 |
| 47 | 09:09:09.284 | +39:34:25.88 | 0.016 | 0.082 | 0.059 | 0.042 | 0.029 |
| 48 | 12:10:22.266 | +16:47:12.45 | 0.032 | 0.166 | 0.119 | 0.085 | 0.059 |
| 51 | 10:00:14.228 | +05:23:03.45 | 0.024 | 0.119 | 0.085 | 0.061 | 0.042 |
| 53 | 10:03:46.445 | +32:53:48.35 | 0.013 | 0.066 | 0.047 | 0.034 | 0.023 |

[17] Based on earlier work by K. D. Gordon et al. (2009), E. L. Fitzpatrick et al. (2019), K. D. Gordon et al. (2021), and M. Decleir et al. (2022).

## ORCID iDs

Alexander Beckett https://orcid.org/0000-0001-7396-3578
Marc Rafelski https://orcid.org/0000-0002-9946-4731
Claudia Scarlata https://orcid.org/0000-0002-9136-8876
Keunho Kim https://orcid.org/0000-0001-6505-0293
Ilias Goovaerts https://orcid.org/0009-0007-8470-5946
Matthew A. Malkan https://orcid.org/0000-0001-6919-1237
Wayne Webb https://orcid.org/0009-0000-9224-8780
Harry Teplitz https://orcid.org/0000-0002-7064-5424
Matthew Hayes https://orcid.org/0000-0001-8587-218X
Vihang Mehta https://orcid.org/0000-0001-7166-6035
Anahita Alavi https://orcid.org/0000-0002-8630-6435
Andrew J. Bunker https://orcid.org/0000-0002-8651-9879
Annalisa Citro https://orcid.org/0009-0000-9676-0538
Nimish Hathi https://orcid.org/0000-0001-6145-5090
Alaina Henry https://orcid.org/0000-0002-6586-4446
Alexandra Le Reste https://orcid.org/0000-0003-1767-6421
Alessia Moretti https://orcid.org/0000-0002-1688-482X
Michael J. Rutkowski https://orcid.org/0000-0001-7016-5220
Maxime Trebitsch https://orcid.org/0000-0002-6849-5375
Anita Zanella https://orcid.org/0000-0001-8600-7008